\documentclass[aps,prd,twocolumns,eqsecnum,preprintnumbers,showpacs,amsmath,amssymb]{revtex4}

\usepackage{graphicx}% Include figure files
\usepackage{dcolumn}% Align table columns on decimal point
\usepackage{bm}% bold math

\begin{document}

\title{ TRUE DYNAMICAL AND GAUGE STRUCTURES OF THE

QCD GROUND STATE AND THE SINGULAR GLUON FIELDS}

\author{Vakhtang Gogokhia}
\email[]{gogohia.vahtang@wigner.hu}
\author{Gergely G\'abor Barnaf\"oldi}
\email[]{barnafoldi.gergely@wigner.hu}

\affiliation{Wigner Research Centre for Physics, 29-33 Konkoly-Thege Mikl\'os Str., H-1121, Budapest, Hungary}

\date{\today}

\begin{abstract}
We convincingly argue that the true dynamical and gauge structure of the QCD ground state is much more complicated than its Lagrangian exact gauge symmetry supposes to be. The dynamical source of these complications has been identified with the tadpole/seagull term, which renormalized version called, the {\sl mass gap}. Its true dynamical role is hidden in the QCD ground state, but it is explicitly present in the full gluon self-energy. To disclose it the splintering between the transverse conditions for the full gluon self-energy and its subtracted counterpart has been derived. The equation of motion for the full gluon propagator on account of the mass gap was given, which allows to fix the dynamical and gauge structures by a newly-introduced generalized gauge. A novel non-perturbative analytical method, the mass gap approach was developed for QCD and its groud state as well.
We have discovered a general, non-perturbative singular solution for the full gluon propagator valid in the whole gluon momentum range, while dominating at large distances over all the other possible solutions. It accommodates all the severe infrared singularities, which may appear in the QCD ground state due to the self-interaction of massless gluon modes. A corresponding non-pertubative multiplicative infrared renormalization program has been formulated. The resulting full gluon propagator prevents gluons to appear at large distances as physical states (confinement of color gluons). Our approach also explains the scale violation in the asymptotic freedom regime, and why the rising potential between heavy quarks is only linear one.
\end{abstract}



\maketitle

\section{Introduction}

\hspace{2mm}

Quantum Chromodynamics (QCD) is widely accepted as the well-functioning quantum field gauge theory of strong interactions, which works not only at the fundamental quark-gluon level, but at the more complex hadronic level as well~\cite{1,2,3,4,5,6,7,69}. This theory should describe the properties of the observed hadrons in terms of the non-observable quarks and gluons from first principles.
In parallel, it is to be complemented by the quark model (QM), which treats the strongly-interacting particles (baryons and mesons)
as bound-states of quarks, emitting and absorbing gluons. This purpose remains a formidable task yet because of the multiple
dynamical and topological complexities of low-energy particle physics, originated from QCD and its ground state.
This happens because QCD as a fundamental theory still suffers from a few conceptual problems. We focus on some of them as follows:

\begin{itemize}
\item[A.] The dynamical generation of a mass squared at the fundamental quark-gluon level, since
the QCD Lagrangian forbids such kind of terms apart from the current quark mass.
\item[B.] Whether the symmetries of the QCD Lagrangian and its ground state coincide or not?
\item[C.] The extension of the mass gap concept to be accounted  for the QCD ground state, since first it has been defined
within the Hamiltonian formalism.
\item[D.] The non-observation of the colored objects as physical states which does not follow from the
QCD Lagrangian, i.e., it cannot explain confinement of gluons and quarks.
\end{itemize}

The properties and symmetries of the QCD Lagrangian, and thus including its Yang-Mills (YM) part, are
well-known~\cite{1,2,3,4,5,6,7,8,69} (and references therein).
As mentioned above, any mass scale parameter apart from the current quark mass, explicitly violates the $SU(3)$ color gauge invariance/symmetry of the QCD Lagrangian, for example such as the massive gluon term $M^2_gA_{\mu}A_{\mu}$. It can be treated as the mass gap, the concept introduced first by
Jaffe and Witten (JW)~\cite{9} within the framework of the QCD/YM Hamiltonian formalism. One of the first aims of our investigation here is to extend the concept of the mass gap to be accounted for the QCD/YM ground state (vacuum) as well. Therefore, we must address to the system of dynamical equations of motion, the so-called Schwinger\,--\,Dyson (SD) system of equations, describing the interactions and propagations of quarks and gluons in the QCD vacuum~\cite{2,8,10,11,12,13,14,15}.
Thus, if there is no room in the QCD Lagrangian  for the mass scale parameter(s), then the only place where it may
explicitly appear is this system, indeed, which contains the full dynamical information on QCD.
In other words, it is not enough to know the Lagrangian of the theory, but it is also necessary and important to know
the true dynamical and gauge structures of its ground state. Furthermore, there might be symmetries of the Lagrangian which
do not coincide with symmetries of the vacuum
and vice versa. These equations should be also complemented by the corresponding Slavnov\,--\,Taylor (ST) identities, which connect lower- and higher Green's functions (propagators and vertices) to each other~\cite{2,8,10,11,12,13,14,15,16,17,18,19,20,21,22,69}. These identities are consequences of the exact gauge invariance and are important for the renormalizability of the QCD. They {\sl "are \ exact \ constraints \ on \ any \ solution \ to \ QCD"}~\cite{2}.
The SD system of dynamical equations, complemented by the ST identities, can serve as an adequate and effective tool for the non-perturbative (NP) analytical
approach to low-energy QCD. Their investigation may reveal much more dynamical information on the QCD ground state, than its Lagrangian may provide at all.
In connection to this let us note that
there exists another powerful NP approach--the lattice QCD--to calculate the properties of low-energy particle physics~\cite{23}.
We believe that these two NP approaches should complement each other, in order to increase our understanding of quantum field theories of particle physics.

We organize our paper as follows: in Section II we introduce the gluon SD equation. In Section III the transversity of the full gluon self-energy
is investigated without use of the perturbation theory (PT). We discuss under which conditions the exact gauge symmetry might be preserved in the QCD ground state. Then we have shown that the true dynamical and gauge structures of the QCD vacuum are not those which
the exact gauge symmetry of its Lagrangian requires. All this made it possible to extend the mass gap concept to be accounted for the the QCD
vacuum as well. In Section IV the mass gap approach to QCD is formulated within a newly-derived generalized gauge. In Sections V and VI
the solution of the full gluon propagator, accommodating all the possible severe infrared (IR) singularities when the gluon momentum goes
to zero, has been found. In Section VII we explain the scale violation in the asymptotic freedom (AF) regime.
In Section VIII the dimensional regularization expansion for the severe IR singularities has been derived. In Section IX
the intrinsically NP (INP) multiplicative (MP) IR renormalization program for the full gluon propagator has been formulated.
In Section X the IR renormalization of the ST identity has been performed. The coincidence of the IR remormalization
constants for the full gluon propagator and the tadpole term has been shown in detail, which is important for the general renormalization
beyond the PT. In Section XI we discuss our results and in Section XII they have been summarized in detail. In Appendix A we demonstrate the
uniqueness of our approach. In Appendix B the linear rising potential between the heavy quarks has been derived.
The $\beta$-function free from all the types of the PT "contaminations" has been derived in Appendix C.
Both results are direct phenomenological consequences of our approach.

\section{The gluon SD equation}
%\label{gluonsd}

The propagation of gluons is one of the main dynamical effects in the QCD vacuum.
The importance of the corresponding equation of motion is due to the fact that its solutions are supposed to reflect the quantum-dynamical structure of the QCD ground state. The gluon SD equation is a highly non-linear (NL) one because of the self-interaction of massless gluon modes, so the number of its independent solutions is not fixed {\sl a priori}. These solutions have to be considered equally. It is important to underline in advance that unlike Quantum
Electrodynamics (QED)~\cite{24}, in QCD any deviation of the full gluon propagator from the free one requires the presence of the mass squared scale parameter
on the general dimensional ground (see below).
The structure of the gluon SD equation is present in this section in some necessary details. For our purpose it is convenient
to begin with the general description of the SD equation for the full gluon propagator $D_{\mu\nu}(q)$. Analytically it can be written down as follows:

\begin{equation}
D_{\mu\nu}(q) = D^0_{\mu\nu}(q) + D^0_{\mu\rho}(q) i \Pi_{\rho\sigma}(q; D) D_{\sigma\nu}(q),
\end{equation}
where $D^0_{\mu\nu}(q)$ denotes the free gluon propagator, while $\Pi_{\rho\sigma}(q; D)$ is the full gluon self-energy which depends on the full gluon propagator due to the non-abelian character of QCD.
Here and everywhere below we omit the color group indices, since for the gluon propagator (and hence for its self-energy) they factorize, for example $D^{ab}_{\mu\nu}(q) = D_{\mu\nu}(q)\delta^{ab}$. The gluon SD eq.~(2.1) in terms of the corresponding skeleton loop diagrams is shown in Fig. 1.

\begin{figure}[h!]
\begin{center}
\includegraphics[width=13.0truecm]{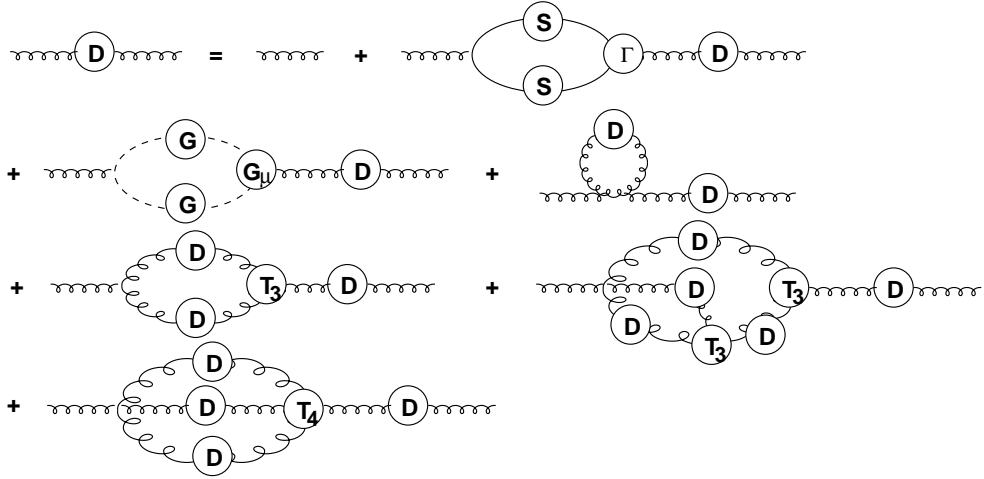}
\caption{The SD equation for the full gluon propagator as present in~\cite{8}.}
\label{fig:1}
\end{center}
\end{figure}

Here helix/stringy line is for the free gluon propagator, while $D$ denotes its full counterpart. The $S$ with solid lines denotes the full quark propagator, and $\Gamma$ denotes the full quark-gluon vertex. $G$ with dashed lines denotes the full ghost propagator, and $G_{\mu}$ is
the full ghost-gluon vertex.
Finally, $T_3$ and $T_4$ denote the full three and four-gluon vertices, respectively.
Fig. 1 shows that the full gluon self-energy is the sum of a few terms, namely

\begin{equation}
\Pi_{\rho\sigma}(q; D) = \Pi^q_{\rho\sigma}(q) + \Pi^{gh}_{\rho\sigma}(q) + \Pi_{\rho\sigma}^t(D) +
\Pi^{(1)}_{\rho\sigma}(q; D^2) + \Pi^{(2)}_{\rho\sigma}(q; D^4) + \Pi^{(2')}_{\rho\sigma}(q; D^3),
\label{2.2}
\end{equation}
where $\Pi^q_{\rho\sigma}(q)$ describes the skeleton loop contribution for the quark degrees of freedom
as an analogue of the vacuum polarization tensor in QED.
Note that here and below the superscript or subscript '$q$' means quark (not to be mixed up with the gluon momentum variable $q$). The $\Pi^{gh}_{\rho\sigma}(q)$ describes the skeleton loop contribution associated with the ghost degrees of freedom. Since neither of the skeleton loop integrals depend on the full gluon propagator $D$, they represent the linear contribution to the gluon SD
equation, and $\Pi_{\rho\sigma}^t(D)$ is the so-called constant skeleton tadpole term. $\Pi^{(1)}_{\rho\sigma}(q; D^2)$ represents the skeleton loop contribution,
containing the triple gluon vertices only. Finally, $\Pi^{(2)}_{\rho\sigma}(q; D^4)$ and $\Pi^{(2')}_{\rho\sigma}(q; D^3)$ describe the skeleton two-loop contributions, which combine the triple and quartic gluon vertices. All these quantities are given by the corresponding skeleton loop diagrams in Fig. 1, and they are independent from each other.
The last four terms explicitly contain the full gluon propagators in the corresponding powers symbolically shown above. They form the NL part of the gluon SD equation. The analytical expressions for the corresponding skeleton loop integrals~\cite{25}, in
which the symmetry and combinatorial coefficients and signs have been included, are not important here. We are not going to calculate any of them
explicitly, and thus to introduce into them any truncations/approximations/assumptions or choose some special gauge.

For further purposes it is convenient to present the full gluon self-energy eq.~(2.2) as follows:

\begin{equation}
\Pi_{\rho\sigma}(q; D) = \Pi^q_{\rho\sigma}(q) + \Pi^{YM}_{\rho\sigma}(q; D) = \Pi^q_{\rho\sigma}(q) + \Pi^g_{\rho\sigma}(q; D) + \Pi_{\rho\sigma}^t(D).
\end{equation}
The explicit expression for the tadpole term is

\begin{equation}
\Pi_{\rho\sigma}^t(D) \sim \int d^4 k D_{\alpha\beta}(k) T^0_{\rho\sigma\alpha\beta} = g_{\rho\sigma} \Delta^2_t(D) =
[T_{\rho\sigma} (q) + L_{\rho\sigma}(q)]\Delta^2_t(D),
\end{equation}
where $L_{\rho\sigma}(q) = q_{\rho} q_{\sigma} / q^2$, so that the tadpole term contributes into the both transverse and longitudinal components of the
full gluon propagator (2.1). The gluon part is the sum of all the other terms in eq.~(2.2), namely

\begin{equation}
\Pi^g_{\rho\sigma}(q; D) = \Pi^{gh}_{\rho\sigma}(q) + \Pi^{(1)}_{\rho\sigma}(q; D^2) +
\Pi^{(2)}_{\rho\sigma}(q; D^4) + \Pi^{(2')}_{\rho\sigma}(q; D^3).
\end{equation}

All the quantities which contribute to the full gluon self-energy eq.~(2.3), and hence eqs.~(2.4)-(2.5), are tensors, having the dimensions of mass squared. All these skeleton loop integrals are therefore quadratically divergent in the PT regime, and so they are assumed to be regularized.
We note, contrary to QED, QCD being a non-abelian gauge theory can suffer from the severe IR singularities in the $q^2 \rightarrow 0$ limit due to the self-interaction of massless gluon modes. Thus, all the possible subtractions at zero may be dangerous \cite{2}. That is why in all the quantities below the dependence on the finite (slightly different from zero) dimensionless subtraction point $\alpha$ is to be understood. In other words, all the subtractions at zero and the Taylor expansions around zero should be understood as the subtractions at $\alpha$ and the structure of the Taylor expansions near $\alpha$, where they are justified to be used. From the technical point of view, however, it is convenient to put formally $\alpha=0$ in all the expressions and derivations below, and to restore the explicit dependence on non-zero $\alpha$ in all the quantities only at the later stage. At the same time, in all the quantities where the dependence on the dimensionless ultraviolet (UV) regulating parameter $\lambda$ and $\alpha$ is not shown explicitly, nevertheless, it should be also
assumed. For example, $\Pi_{\rho\sigma}(q; D) \equiv \Pi_{\rho\sigma}(q; D, \lambda, \alpha)$ and similarly for all other quantities. This means that all the expressions are regularized (they become finite), and thus a mathematical meaning is assigned to all of them. In this connection, let us underline that the tadpole term (2.4) is quadratically UV divergent constant $\Delta^2_t(D)$, but already regularized one from below and above as well as all the other
such kind of constants which will appear in what follows. Within our approach nothing will depend on how exactly these regulating parameters have been introduced. They will disappear from the theory after the NP renormalization program will be performed. For more detailed description of the general
structure and properties of the SD system of equations see the above-cited references. Let us also remind that the whole gluon momentum range
is $q^2 \in [0, \infty)$. In what follows we will work in Euclidean metric $q^2=q^2_0 + \bf q^2$ since it implies $q_i \rightarrow 0$ when $q^2 \rightarrow 0$
and vice versa. This makes it possible to avoid the un-physical IR singularities on the light cone (see Section VIII).

\section{Transversity of the full gluon self-energy}
%\label{fullgtrans}

The first step in the renormalization program of any gauge theory is the removal of the UV quadratic divergences (if any) in order to make
the corresponding theory renormalizable. This can be achieved by introducing suitable subtraction scheme in order to separate them from the PT logarithmic divergences. In this connection it is worth mentioning that a preliminary step in this program, namely to regularize our expressions, has been already done by introducing the corresponding regulating parameters $\lambda$ and $\alpha$ in the previous Section II. They symbolize that the regularization can be performed by any means, but how exactly is not important, as underlined above.

The basic relation to which the full gluon propagator should satisfy is the corresponding ST identity, namely
\begin{equation}
q_{\mu}q_{\nu} D_{\mu\nu}(q) = i \xi,
\end{equation}
where $\xi$ is the gauge-fixing parameter.
It is a consequence of the color gauge invariance/symmetry of QCD and, as emphasized above, is an exact constraint on any solution  to QCD.
The ST identity (3.1) implies that the general tensor decomposition of the full gluon propagator in the covariant gauge is as follows:
\begin{equation}
D_{\mu\nu}(q) = i \left[ T_{\mu\nu}(q) d(q^2) + \xi L_{\mu\nu}(q) \right] {1 \over q^2},
\end{equation}
where the invariant function $d(q^2)= d(q^2; \xi)$ is the corresponding Lorentz structure of the full gluon propagator (further - the gluon invariant function,
for simplicity). Also, throughout this paper we use the standard definition of $T_{\mu\nu}(q) = \delta_{\mu\nu} - q_{\mu}q_{\nu} / q^2= \delta_{\mu\nu} - L_{\mu\nu}(q)$ in Euclidean metric.

If one neglects all the contributions to the full gluon self-energy in eq.~(2.1), i.e., putting formally $d(q^2)=1$ in eq.~(3.2), then one obtains the free gluon propagator, namely
\begin{equation}
D^0_{\mu\nu}(q) = i \left[ T_{\mu\nu}(q) + \xi_0 L_{\mu\nu}(q) \right] (1 / q^2),
\end{equation}
where $\xi_0$ is the corresponding gauge-fixing parameter. The general ST identity (3.1) will look like
\begin{equation}
q_{\mu}q_{\nu} D^0_{\mu\nu}(q) = i \xi_0.
\end{equation}

Contracting now the full gluon SD eq.~(2.1) with $q_{\mu}$ and $q_{\nu}$, and doing some algebra on account of the previous relations (3.1)-(3.4),
one finally obtains

\begin{equation}
q_{\rho}q_{\sigma} \Pi_{\rho\sigma}(q; D) = {( \xi_0 - \xi) \over \xi \xi_0 } (q^2)^2.
\end{equation}
From this relation it follows that by itself it cannot remove the quadratic UV divergences from the theory.  As pointed out above, we will achieve this by formulating the suitable subtraction scheme in which the corresponding transverse conditions can be implemented. But whether they will be satisfied (i.e., equal zero) or not requires much more careful investigation in QCD.

Let us start the formulation of the subtraction scheme for the full gluon self-energy eq.~(2.3) as follows:
\begin{equation}
\Pi^{(s)}_{\rho\sigma}(q; D)= \Pi_{\rho\sigma}(q; D) - \Pi_{\rho\sigma}(0; D)= \Pi_{\rho\sigma}(q; D) - \delta_{\rho\sigma}\Delta^2(D),
\end{equation}
and thus $\Pi^{(s)}_{\rho\sigma}(0; D) = 0$, by definition, and where
\begin{eqnarray}
\Pi_{\rho\sigma}(0)  &=& \Pi^q_{\rho\sigma}(0) + \Pi^g_{\rho\sigma}(0; D) + \Pi^t_{\rho\sigma}(D) \nonumber\\
&=& \delta_{\rho\sigma}\Delta^2(D) = \delta_{\rho\sigma}[\Delta^2_q + \Delta^2_g(D) + \Delta^2_t(D)]
\end{eqnarray}
is the sum of the corresponding skeleton loop integrals at $q=0$ contributing to eq.~(2.3), while $\Delta^2_g(D)$ itself is the sum of the
corresponding skeleton loop integrals at $q=0$ contributing to eq.~(2.5).
All of them are quadratically UV divergent, but already regularized constants.
Let us remind that the subtraction at zero is to be understood in a such way that we subtract at $q^2 = - \mu^2$ with $\mu^2 \rightarrow 0$ final limit. It is worth noting that the subtraction (3.6) is equivalent to add zero to the corresponding identity. Indeed,
$\Pi_{\rho\sigma}(q; D)= \Pi_{\rho\sigma}(q; D) - \Pi_{\rho\sigma}(0; D) + \Pi_{\rho\sigma}(0; D)$ and denoting
$\Pi^{(s)}_{\rho\sigma}(q; D)= \Pi_{\rho\sigma}(q; D) - \Pi_{\rho\sigma}(0; D)$, one obtains eq.~(3.6). It means that our subtraction scheme change nothing in the initial skeleton loop expressions.

Contracting now eq.~(3.6) with $q_\rho$ and $q_\sigma$, one obtains

\begin{equation}
q_{\rho}q_{\sigma} \Pi^{(s)}_{\rho\sigma}(q; D) = {( \xi_0 - \xi) \over \xi \xi_0 } (q^2)^2 - q^2 \Delta^2(D).
\end{equation}

The general decompositions of the gluon self-energy and its subtracted counterpart into the independent tensor structures look like
\begin{eqnarray}
\Pi_{\rho\sigma}(q) &=& T_{\rho\sigma}(q) q^2 \Pi_t(q^2) - q_{\rho} q_{\sigma} \Pi_l(q^2), \nonumber\\
\Pi^{(s)}_{\rho\sigma}(q) &=& T_{\rho\sigma}(q) q^2 \Pi^{(s)}_t(q^2) - q_{\rho} q_{\sigma} \Pi^{(s)}_l(q^2),
\end{eqnarray}
where in all the quantities the dependence on $D$ is omitted, for simplicity, and will be restored below when necessary.
Here and everywhere below all the invariant functions are dimensionless ones of their argument $q^2$: otherwise they remain arbitrary. However,
both invariant functions $\Pi^{(s)}_t(q^2)$ and $\Pi^{(s)}_l(q^2)$ cannot have power-type singularities (or, equivalently, pole-type ones) at small $q^2$, since $\Pi^{(s)}_{\rho\sigma}(0) =0$ by definition in eq.~(3.6).
Thus, one has the two transverse conditions (3.5) and (3.8) for the four invariant functions, which appear in the decompositions (3.9).

Substituting both decompositions into the subtraction (3.6), and doing some algebra on account of the transverse conditions (3.5) and (3.8),
one arrives at
\begin{eqnarray}
\Pi^{(s)}_t (q^2) &=&  \Pi_t(q^2) - {\Delta^2(D) \over q^2}, \nonumber\\
\Pi^{(s)}_l(q^2) &=& \Pi_l(q^2) + {\Delta^2(D) \over q^2} = - {( \xi_0 - \xi) \over \xi \xi_0 } + {\Delta^2(D) \over q^2},
\end{eqnarray}
where the invariant function $\Pi^{(s)}_t (q^2)$ may still have the logarithmic divergences only in the PT, since all the quadratic UV divergences
summarized into the scale parameter $\Delta^2(D)$, have been already subtracted from the initial invariant function $\Pi_t(q^2)$.
Then for the full gluon self-energy one gets
\begin{equation}
\Pi_{\rho\sigma}(q) = T_{\rho\sigma}(q) \left[ q^2 \Pi^{(s)}_t(q^2) + \Delta^2(D) \right] + L_{\rho\sigma}{( \xi_0 - \xi) \over \xi \xi_0 } q^2
\end{equation}
and substituting it into the gluon SD eq.~(2.1), one obtains
\begin{eqnarray}
D_{\mu\nu}(q) = D^0_{\mu\nu}(q) &+& D^0_{\mu\rho}(q)i T_{\rho\sigma}(q) \left[ q^2 \Pi^{(s)}_t(q^2) +  \Delta^2(D) \right] D_{\sigma\nu}(q) \nonumber\\
&+& D^0_{\mu\rho}(q)i L_{\rho\sigma}(q) {( \xi_0 - \xi) \over \xi \xi_0 } q^2 D_{\sigma\nu}(q),
\end{eqnarray}
which along with the general transverse conditions (3.5) and (3.8) represent the system of equations for the full gluon propagator,
explicitly depending on the mass scale parameter $\Delta^2(D)$.

However, concluding let us underline that contracting it with  $q_{\mu}$ and $q_{\nu}$, one obtains identities $\xi =\xi$ and $\xi_0 = \xi_0$, and
not $\xi$ as a function of $\xi_0$, i.e., $\xi = f(\xi_0)$. In fact, the expression (3.12) is not an equation, but it is an identity! Thus, the transverse relations (3.5) and (3.8) failed to find the function $\xi = f(\xi_0)$ and, at the same time, to change the nature of the expression (3.12).
The important conclusion then is as follows: the only way to get out of these troubles (getting us nowhere) is to satisfy (i.e., put zero) at
least one of them. Only this will
make from the expression (3.12) an equation for the full gluon propagator, and thus to fix the function $\xi = f(\xi_0)$ as well (see below).

\subsection*{Preservation of the exact gauge symmetry and the role of the ghost term}

Let us now show in detail how the $SU(3)$ color gauge symmetry of the QCD Lagrangian might be preserved/saved in its ground state.
By the substitution of the general decompositions (3.2) and (3.3) into eq.~(3.12) and doing some algebra, one obtains

\begin{equation}
D_{\mu\nu}(q) = i \left[ { 1 \over 1 + \Pi^{(s)}_t(q^2) +  \Delta^2(D) / q^2 } T_{\mu\nu}(q)  + \xi L_{\mu\nu}(q) \right] {1 \over q^2},
\end{equation}
i.e., we express the gluon invariant function $d(q^2)$ in terms of $\Pi^{(s)}_t(q^2)$ and $\Delta^2(D)$ but the gauge-fixing parameter
$\xi$ remains the same. However, if it is not changed, i.e., is not determined/fixed as a function of $\xi_0$, then from the second of
the relations (3.10) it follows that the mass scale parameter
$\Delta^2(D)$ should be disregarded on a general ground, since the invariant function $\Pi^{(s)}_l(q^2)$ cannot have the
pole-types singularities, by definition. This means that the UV divergent, but already regularized constant $\Delta^2(D)$ is to be put
zero everywhere $\Delta^2(D)=0$. Then from eq.~(3.7) it follows

\begin{equation}
\Delta^2_q = \Delta^2_g(D) = \Delta^2_t(D) =0,
\end{equation}
i.e., each of these constants should be omitted in the theory, since they are independent from each other (in accordance with the decomposition (2.3)).
This means that the both transverse conditions for the full gluon self-energy (3.5) and its subtracted counterpart (3.8) are
equal to each other, namely

\begin{equation}
q_{\rho}q_{\sigma} \Pi_{\rho\sigma}(q; D) = q_{\rho}q_{\sigma} \Pi^{(s)}_{\rho\sigma}(q; D) = {( \xi_0 - \xi) \over \xi \xi_0 } (q^2)^2.
\end{equation}
The gluon SD eq.~(3.12) becomes

\begin{eqnarray}
D_{\mu\nu}(q) = D^0_{\mu\nu}(q) &+& D^0_{\mu\rho}(q)i T_{\rho\sigma}(q) q^2 \Pi^{(s)}_t(q^2) D_{\sigma\nu}(q) \nonumber\\
&+& D^0_{\mu\rho}(q)i L_{\rho\sigma}(q) {( \xi_0 - \xi) \over \xi \xi_0 } q^2 D_{\sigma\nu}(q),
\end{eqnarray}
while its 'solution' looks like

\begin{equation}
D_{\mu\nu}(q) = i \left[ { 1 \over 1 + \Pi^{(s)}_t(q^2)} T_{\mu\nu}(q)  + \xi L_{\mu\nu}(q) \right] {1 \over q^2}.
\end{equation}
The ST identities (3.1) and (3.4) are respected, of course. It is interesting to note that all the quadratically divergent but regularized constants have to disappear from the theory because the corresponding invariant function cannot have the pole-types singularities, while
the corresponding transverse conditions (3.15) are not yet satisfied.

Therefore the relation (3.16) still remains an identity than an equation. As underlined above, to make it equation, the one or both transverse
conditions for the full gluon self-energy and its subtracted counterpart should be satisfied. So that in this case from the relations (3.15), one obtains

\begin{equation}
q_{\rho}q_{\sigma} \Pi_{\rho\sigma}(q; D^{PT}) = q_{\rho}q_{\sigma} \Pi^{(s)}_{\rho\sigma}(q; D^{PT}) =0, \quad \textrm{which requires} \quad \xi =\xi_0,
\end{equation}
and vice versa, i.e., if $\xi =\xi_0$ then the relations (3.15) have to be satisfied.

The relation (3.16) becomes equation now, namely

\begin{equation}
D^{PT}_{\mu\nu}(q) = D^0_{\mu\nu}(q) + D^0_{\mu\rho}(q)i T_{\rho\sigma}(q) q^2 \Pi^{(s)}_t(q^2) D^{PT}_{\sigma\nu}(q),
\end{equation}
because from it one arrives at

\begin{equation}
q_{\mu} q_{\nu}(q)D^{PT}_{\mu\nu}(q) = q_{\mu} q_{\nu}D^0_{\mu\nu}(q)  = i \xi_0,
\end{equation}
and thus the gauge-fixing parameter of the PT full gluon propagator is fixed as follows: $\xi = f(\xi_0)= \xi_0$, in complete agreement with eq.~(3.18).
Its 'solution' (3.17) now looks like

\begin{equation}
D^{PT}_{\mu\nu}(q) = i \left[ d^{PT}(q^2) T_{\mu\nu}(q)  + \xi_0 L_{\mu\nu}(q) \right] {1 \over q^2}, \quad  d^{PT}(q^2) = { 1 \over 1 + \Pi^{(s)}_t(q^2)},
\end{equation}
where the invariant function $\Pi^{(s)}_t(q^2)=\Pi^{(s)}_t(q^2; D^{PT})$ is regular function of its argument, free from the quadratic UV divergences, but still may have only the logarithmic ones in the PT $q^2 \rightarrow \infty$ limit. Obviously, this equation describes the propagation of the PT massless gluons, since it has the PT singularity on the mass-shell $q^2=0$ only. For the explanation of the notation of the full gluon propagator as $D^{PT}$ in this case see concluding remarks in this subsection. Let us note in advance that the equality (3.20) takes place only for the regularized massless gluon fields, for their renormalized counterparts these gauge-fixing parameters should be different, in principle. However, within the INP singular solution
for the full gluon propagator derived in this work this difference is not important (see Sections IV-X).

The absence of the mass scale parameters in the system of eqs.~(3.18)-(3.21) can be now attributed to the satisfied transverse conditions
(3.18) as well. Combining it with the second of the decompositions (3.9), from the second of the relations (3.10), one arrives at
$q^2 \Pi^{(s)}_l(q^2) = \Delta^2(D) = 0$ when $\xi =\xi_0$, which also implies eq.~(3.14). It is possible to say that just
the satisfied transverse conditions (3.18) decrease the quadratic UV divergences of the corresponding skeleton loop integrals to a logarithmic ones at large $q^2$. In other words, due to the transverse relations (3.18) all the mass scale parameters, having dimensions of mass squared, shown in eq.~(3.14), should be removed/disappeared from the theory, i.e., there is no the explicit presence of such kind of parameters in the PT gluon SD equation and thus in the PT gluon propagator as well. For how precisely the satisfied transverse relations work in order to remove from the theory the quadratically divergent, but regularized constants, see next subsection.

Let us now make a few remarks concerning the role of the ghost term in the satisfied transverse condition (3.18). From the relation (2.3) and in the absence of the tadpole term, its YM part is reduced to the gluon part, defined in eq.~(2.5). It is well-known that in QCD the quark contribution in the relation (2.3) can be made transverse independently from its YM part, and thus the quark constant $\Delta^2_q$ is to be finally removed from the theory.
Then the satisfied transverse condition (3.18) is reduced to
$q_{\rho} q_{\sigma} \Pi^g_{\rho\sigma}(q; D) = q_{\rho} q_{\sigma} [ \Pi^{gh}_{\rho\sigma}(q) + \Pi^{(1)}_{\rho\sigma}(q; D^2)
+ \Pi^{(2)}_{\rho\sigma}(q; D^4) + \Pi^{(2')}_{\rho\sigma}(q; D^3)] = 0$ and thus the gluon constant $\Delta^2_g(D)$ is to be finally removed from the theory
as well. Let us note in advance that these constants have to be finally omitted in any case, i.e., not only in the PT QCD (see derivations in the next subsection). It is worth reminding now that none of these terms can satisfy the above-displaced transverse condition separately from each other.
In other words, it cannot be re-written down as the sum of the satisfied transverse relations for each term.
 The role of ghost degrees of freedom is to cancel the un-physical (longitudinal) component of the full gluon propagator. Therefore this transverse condition
is important for ghosts to fulfill their role, and thus to maintain the unitary of the $S$-matrix in the PT QCD. Just the Faddeev\,--\,Popov ghost contribution, $\Pi^{gh}_{\rho\sigma}(q)$ makes this transverse relation valid.
For the explicit demonstration of how the ghosts guarantee this transverse condition in lower orders of the PT see, for example~\cite{3,4,5}. This should be true in every order of the PT in agreement with the above-mentioned transverse condition where the skeleton gluon loop diagrams are present.
However, from our analysis above it follows that ghost term makes the transverse conditions (3.18) valid if and only if $\xi=\xi_0$. Note, in the
derivation of the transverse relations (3.18) we did not specify the content of the gluon part of the full gluon self-energy, shown in eq.~(2.5).
Therefore the equality $\xi=\xi_0$ is the first
necessary condition, while the presence of the ghost term there is the second sufficient one for the transverse conditions
(3.18) to be valid. Both conditions are important in the PT QCD, of course. However, in the NP QCD, which is main subject of our investigation in this paper,
the equality $\xi=\xi_0$ will not take the place, i.e, will be violated, while the ghost term will be retained in the full gluon self-energy.
Within our approach to NP QCD the transversity of the full gluon propagator will be achieved in the completely different way, not depending on the
presence of the ghost term. However, it will come into the play again in the PT $q^2 \rightarrow \infty$ limit, which will lead to $\xi=\xi_0$, as expected.

The system of eqs.(3.18)-(3.21) is free of all the types of the scale parameters having the dimensions of mass squared, forbidden by the exact
gauge symmetry of the QCD Lagrangian. Therefore, it constitutes that the gauge symmetry of the QCD ground state, reflected by this system,
coincides with the symmetry of its Lagrangian. As it was described in \cite{8}, such coincidence
is analogous to QED where the abelian $U(1)$ gauge symmetry of the Lagrangian is the same as of its ground state. That is why we denoted the full gluon propagator in this system as $D^{PT}$ (suppressing the term $\Delta^2(D) / q^2$ in the PT  $q^2 \rightarrow \infty$ limit in the expression (3.13), one arrives at the expression (3.21)). The essential feature of this phenomenon is the equality $\xi = \xi_0$.
In this connection, let us note that one can start from the expression (3.2). It is nothing else but the general decomposition of the tensor function (the full gluon propagator), depending on the one variable, into the independent tensor structures.
Then it will automatically satisfy to the relation (3.1), called the ST identity and treated the gauge-fixing parameter as a some function of $q^2$, i.e.,
$\xi = f(q^2)$. This has nothing to do with the statement that the ST identity is a consequence of the exact gauge symmetry. It becomes of its consequence, indeed, when the gauge-fixing function is fixed by the relation (3.20), as it has been described in this subsection within the PT methodology.
However, unlike QED,  QCD/YM, being a non-abelian gauge theory with the strong coupling constant, is to be treated beyond the PT, especially its ground state.
So that the general case when $\xi \neq \xi_0$ should be investigated in detail, while keeping the ST identities (3.1) and (3.3) valid, anyway.

\subsection*{Violation of the exact gauge symmetry and the role of the tadpole term}

The distinctive feature of the gluon SD eq.~(3.19) is the absence of any mass squared scale parameters in its dynamical and gauge structures.
None such type of parameters is present in its transverse component, reflecting the corresponding dynamics, nor in its longitudinal
counterpart, reflecting the gauge choice. However, such a scale parameter, namely the tadpole term, is explicitly present in the initial
gluon SD eq.~(2.1) and see Fig. 1 as well. By the virtue of the exact gauge symmetry
all the scale parameters should be disregarded on a general ground. It is important to note that the quark and gluon scale
parameters  $\Delta^2_q$ and $\Delta^2_g(D)$, respectively, are not explicitly present in the full gluon self-energy, but appear as a result of the
corresponding gauge-invariant subtraction scheme. At the same time, the tadpole term $\Delta^2_t(D)$ is explicitly present in the full
gluon-self-energy from the very beginning, making thus the direct contributions into the dynamical and gauge structures of the QCD vacuum.
If it has to be removed along with other quadratically divergent but regularized constants
due to the gauge invariance, then a natural question arises why is it present in the gluon SD eq.~(2.1) at all? Especially knowing that it makes the YM theory explicitly unrenormalizable! in the PT sense and preventing the ghosts to cancel the longitudinal component of the full gluon propagator and thus making it not transverse. So one may conclude that the system of eqs.~(3.18)-(3.21) hides the role of the tadpole term
in the QCD ground state, making it "invisible".
Nothing would have been changed in the derivation of the above-mentioned system of equations, if the tadpole term in the gluon SD eq.~(2.1)
were not existed, indeed. In other words, it plays no any dynamical role in the preservation of the exact gauge symmetry in the QCD ground state
(one cannot impose any condition of the cancelation of the quark and gluon scale parameters  $\Delta^2_q$ and $\Delta^2_g(D)$ by $\Delta^2_t(D)$).
Let us emphasize that we are discussing the presence/existence of the tadpole term in the vacuum of QCD from the
dynamical ('physical') point of view, but not its combinatorial (mathematical) meaning.

Just in order to disclose/reveal the true role of the tadpole term in the dynamical and gauge structures of the QCD ground state,
let us investigate the subtraction scheme (3.6) for the full gluon self-energy in more detail.
For this purpose it is instructive to remind the initial eq.~(2.3), which is

\begin{equation}
\Pi_{\rho\sigma}(q; D) = \Pi^q_{\rho\sigma}(q) + \Pi^g_{\rho\sigma}(q; D) + \delta_{\rho\sigma} \Delta^2_t(D),
\end{equation}
and therefore

\begin{equation}
\Pi_{\rho\sigma}(0; D) = \Pi^q_{\rho\sigma}(0) + \Pi^g_{\rho\sigma}(0; D) + \delta_{\rho\sigma} \Delta^2_t(D).
\end{equation}

The initial subtraction scheme (3.6) holds, namely

\begin{equation}
\Pi^s_{\rho\sigma}(q; D) = \Pi_{\rho\sigma}(q; D) - \Pi_{\rho\sigma}(0;D),
\end{equation}
where

\begin{equation}
\Pi^s_{\rho\sigma}(q; D) = [\Pi^q_{\rho\sigma}(q) - \Pi^q_{\rho\sigma}(0)] + [\Pi^g_{\rho\sigma}(q; D) - \Pi^g_{\rho\sigma}(0;D)] = \Pi^{q(s)}_{\rho\sigma}(q)
+ \Pi^{g(s)}_{\rho\sigma}(q; D),
\end{equation}
so that $\Pi^{q(s)}_{\rho\sigma}(0) = \Pi^{g(s)}_{\rho\sigma}(0; D) =0$, by definition, and thus $\Pi^s_{\rho\sigma}(0; D)=0$ as well.
The subtraction (3.24), on account of the relations (3.23), can be expressed as follows:

\begin{equation}
\tilde \Pi^{(s)}_{\rho\sigma}(q; D) = \Pi_{\rho\sigma}(q; D) - \delta_{\rho\sigma}\Delta^2_t(D)
\end{equation}
because of the relations (3.7), and where

\begin{equation}
\tilde \Pi^{(s)}_{\rho\sigma}(q; D) = [\Pi^{q(s)}_{\rho\sigma}(q) + \delta_{\rho\sigma} \Delta^2_q] + [\Pi^{g(s)}_{\rho\sigma}(q; D) + \delta_{\rho\sigma} \Delta^2_g(D)].
\end{equation}
In the initial subtraction (3.6) we did not specify the context of the subtracted gluon self-energy,
while in the both subtracted relations (3.24) and (3.26) we did this and they were present in the relations (3.25) and (3.27), respectively.
From these relations it is explicitly seen that the detailed subtracted parts of the
full gluon self-energy are free from the tadpole term $\Delta^2_t(D)$, as expected.
The difference between these two subtracted gluon self-energies is

\begin{equation}
\tilde \Pi^{(s)}_{\rho\sigma}(q; D) - \Pi^{(s)}_{\rho\sigma}(q; D) = \delta_{\rho\sigma}[\Delta_q^2 + \Delta_g^2(D)].
\end{equation}
From this relation one concludes that contrary to $\Pi^{(s)}_{\rho\sigma}(0; D)=0$, by definition, the auxiliary subtracted term
$\tilde \Pi^{(s)}_{\rho\sigma}(0; D) = \delta_{\rho\sigma}[\Delta_q^2 + \Delta_g^2(D)]$, i.e., it is non-zero constant at this stage, which becomes
zero at final stage only.

Our final goal in this subsection is to find such a gluon SD equation which retain the tadpole term in its structure, but will be free from the quark and gluon constants, since the former one is explicitly present in the initial gluon SD eq.~(2.1), while the latter ones not. On the other hand, we already know that for this purpose one of the transverse conditions should be satisfied, i.e., put zero in order to transform the expression (3.12) into the equation for full gluon propagator. In order to distinguish between the constants $\Delta_q^2$, $\Delta_g^2(D)$ and the tadpole term $\Delta_t^2(D)$,
let us begin with the auxiliary/spurious (or, equivalently, detailed) subtracted gluon self-energy (3.27). It explicitly depends on the quark and gluon
constants only, so that

\begin{equation}
q_{\rho} q_{\sigma} \tilde \Pi^{(s)}_{\rho\sigma}(q; D) = 0.
\end{equation}
It can be reduced to the two independent satisfied transverse conditions due to the sum (3.27) as follows:

\begin{eqnarray}
q_{\rho} q_{\sigma} [\Pi^{q(s)}_{\rho\sigma}(q) + \delta_{\rho\sigma} \Delta^2_q] &=& 0, \nonumber\\
q_{\rho} q_{\sigma} [\Pi^{g(s)}_{\rho\sigma}(q; D) + \delta_{\rho\sigma} \Delta^2_g(D)] &=& 0.
\end{eqnarray}
It is worth reminding from the previous subsection that such separation of the satisfied transverse conditions, shown in eqs.~(3.30),
is possible in QCD. The quark contribution can be
made transverse (the first of the relations (3.30)) independently from its gluon part (the second of the relations (3.30)), owing to the fact that the current which flows around the closed skeleton quark loop (see the first skeleton loop diagram in Fig. 1) is conserved from the very beginning. This is in complete analogy with QED, where the current flowing around the closed skeleton electron loop (vacuum polarization tensor, which only one contributes into the full photon self-energy) is conserved.

Substituting the corresponding independent tensor decompositions

\begin{eqnarray}
\Pi^{q(s)}_{\rho\sigma}(q) &=&  T_{\rho\sigma}(q) q^2 \Pi^{q(s)}_t(q^2) - q_{\rho} q_{\sigma} \Pi^{q(s)}_l(q^2), \nonumber\\
\Pi^{g(s)}_{\rho\sigma}(q; D) &=&  T_{\rho\sigma}(q) q^2 \Pi^{g(s)}_t(q^2;D) - q_{\rho} q_{\sigma} \Pi^{g(s)}_l(q^2; D)
\end{eqnarray}
into the transverse relations (3.30), and doing some algebra, one obtains

\begin{equation}
\Pi^{q(s)}_l(q^2) =  { \Delta^2_q \over q^2}, \quad \quad  \Pi^{g(s)}_l(q^2; D)=  { \Delta^2_g(D) \over q^2}.
\end{equation}
However, both relations are not possible, since $\Pi^{q(s)}_l(q^2)$ and  $\Pi^{g(s)}_l(q^2;D)$
by themselves cannot have power-type singularities at small $q^2$, because of the relations $\Pi^{q(s)}_{\rho\sigma}(0) = \Pi^{g(s)}_{\rho\sigma}(0; D)=0$,
displayed above.
This means that the quadratically UV divergent, but already regularized constant $\Delta^2_q$ and $\Delta^2_g(D)$ are to be disregarded on a general ground, i.e. put zero everywhere

\begin{equation}
\Delta^2_q = \Delta^2_g(D)=0, \quad \textrm{which means that} \quad \Delta^2 (D) = \Delta^2_t(D) \neq 0,
\end{equation}
as it comes out from the relation (3.7), while the tadpole contribution $\Delta^2_t(D)$ remains intact. From the relations (3.32)
and (3.33) one concludes that $\Pi^{q(s)}_l(q^2) = \Pi^{g(s)}_l(q^2; D)= 0$ as well, which means the satisfied transverse relations
$q_{\rho} q_{\sigma}\Pi^{q(s)}_{\rho\sigma}(q) = q_{\rho} q_{\sigma} \Pi^{g(s)}_{\rho\sigma}(q; D) = 0$ take place due to the relations (3.31).
Then from the relations (3.25) it follows that the initial $\Pi^{(s)}_{\rho\sigma}(q)$, defined in eq.~(3.6), is also transverse, i.e.,

\begin{equation}
q_{\rho} q_{\sigma} \Pi^{(s)}_{\rho\sigma}(q; D) = 0.
\end{equation}
From it, and due to the relations (3.9), (3.10) and (3.33), one finally obtains

\begin{eqnarray}
\Pi^{(s)}_t (q^2) &=&  \Pi_t(q^2) - {\Delta_t^2(D) \over q^2}, \nonumber\\
\Pi^{(s)}_l(q^2) &=& - {( \xi_0 - \xi) \over \xi \xi_0 } + {\Delta_t^2(D) \over q^2} = 0.
\end{eqnarray}

In connection with the second of the relations (3.35), let us remind that it can be satisfied if $\xi = \xi_0$, then $\Delta^2_t(D)=0$ as well, since the invariant function $\Pi^{(s)}_l(q^2)$ cannot have the pole-type singularities, by definition. Just these conditions have been used in order to get to the system of eqs.~(3.18)-(3.21), which preserves the exact gauge symmetry of the QCD Lagrangian in its ground state. However, it has another
(more general in our opinion)
solution. It shows the gauge change ($\xi$ is different from $\xi_0$), so that the gauge-fixing parameter for the full gluon propagator $\xi$ becomes the corresponding function of $q^2$, i.e., $\xi= \xi(q^2)$. From the second equation in the relations (3.35), one gets

\begin{equation}
{( \xi_0 - \xi) \over \xi \xi_0 } = {\Delta_t^2(D) \over q^2},
\end{equation}
precisely which solution will determine the above-mentioned function $\xi= \xi(q^2)$. Its explicit solution
is convenient to postpone until the next section. Using it in the transverse condition (3.5), one arrives at the two independent transverse conditions as follows:

\begin{eqnarray}
q_{\rho}q_{\sigma} \Pi_{\rho\sigma}(q; D) &= &  q^2 \Delta^2_t(D)  \neq 0,  \nonumber\\
q_{\rho}q_{\sigma} \Pi^{(s)}_{\rho\sigma}(q; D) &=& 0,
\end{eqnarray}
instead of the transverse conditions (3.18). That is why we call such system of the transverse conditions as a result of the splintering procedure, since by formally neglecting the tadpole term they become the same, i.e., reduced to the relations (3.18), and thus there is no need in their separate treatment.

The initial gluon SD eq.~(2.1), via the auxiliary eq.~(3.12), now becomes

\begin{eqnarray}
D_{\mu\nu}(q) = D^0_{\mu\nu}(q) &+& D^0_{\mu\rho}(q)i T_{\rho\sigma}(q) \left[ q^2 \Pi^{(s)}_t(q^2) +  \Delta^2_t(D) \right] D_{\sigma\nu}(q) \nonumber\\
&+& D^0_{\mu\rho}(q)i L_{\rho\sigma}(q) \Delta^2_t(D) D_{\sigma\nu}(q),
\end{eqnarray}
on account of the relations (3.33) and (3.36). The corresponding ST identities therefore are not equal to each other

\begin{equation}
q_{\mu} q_{\nu}(q)D_{\mu\nu}(q) \neq q_{\mu} q_{\nu}D^0_{\mu\nu}(q), \quad \textrm{which implies} \quad \xi \neq \xi_0,
\end{equation}
and vice versa, i.e., when $\xi \neq \xi_0$ then an equality is impossible.

If the tadpole term is formally omitted in the system of eqs.~(3.37)-(3.39), then it will be reduced to the system of eqs.~(3.18)-(3.21), as pointed
out just above.
So that the system of eqs.~(3.18)-(3.21), preserving the gauge symmetry of the QCD Lagrangian in its ground state, is a particular case
of the general system of eqs.~(3.37)-(3.39), which reflects the violation of the Lagrangian gauge symmetry in its ground state. It is important to
understand that the system of eqs.(3.18)-(3.21) describes a hypothetical situation in QCD, while the system of eqs.~(3.37)-(3.39) describes the real
dynamical and gauge structures of its ground state. The solutions of the system of eqs.~(3.18)-(3.21) can be part of the solutions of the system of eqs.~(3.37)-(3.39) in the sense that the latter one also posses the PT $q^2\rightarrow \infty$ limit, when the contribution $\Delta_t(D) / q^2$ can be neglected. However, the final results will be different, leading to the AF regime for the full gluon propagator (3.38), while the gluon SD eq.~(3.19)
cannot provide such regime (see below).

The final system of eqs.(3.37)-(3.39) will not depend on how precisely one introduces the subtraction scheme.
For example, one can define the subtracted gluon self-energy, instead of (3.27), as follows: $\bar \Pi^{(s)}_{\rho\sigma}(q; D)= \Pi^q_{\rho\sigma}(q) + \Pi^g_{\rho\sigma}(q; D)$ in the initial expression (3.22). Note, this definition also does not  depend on the tadpole term, as requested, and $\bar \Pi^{(s)}_{\rho\sigma}(0; D) =\delta_{\rho\sigma}[\Delta_q^2 + \Delta_g^2(D)]$ as well. But from eq.~(3.27) it follows that $\bar \Pi^{(s)}_{\rho\sigma}(q; D)= \tilde \Pi^{(s)}_{\rho\sigma}(q; D)$. Repeating the derivations, it is easy to show that
$q_{\rho}q_{\sigma} \bar \Pi^{(s)}_{\rho\sigma}(q; D) = q_{\rho}q_{\sigma} \tilde \Pi^{(s)}_{\rho\sigma}(q; D) = 0$.
The uniqueness of our approach (in order to establish
the true dynamical and gauge structures of the QCD ground state) will be demonstrated from the general point of view in Appendix A.
The uxiliary/spurious scheme (3.26) described above makes it possible to distinguish between the quark and gluon constants, which are not present in the initial gluon SD eq.~(2.1), and the tadpole constant, which is explicitly present in it. The initial subtraction scheme (3.6) failed to do this. At the same time,
the detailed subtraction scheme finally leads to the satisfied transverse relation (3.34) for the initially subtracted gluon self-energy (3.24), as expected.

Concluding, it is important to emphasize once more that we are under an obligation to satisfy the transverse relation for the subtracted gluon self-energy
(defined in any possible way). Reminding that this is necessary to do in order to make from the expression (3.12) an equation for the full gluon propagator.
In other words, none of the above-mentioned satisfied transverse relations have been introduced by hand, but they have been implemented
into the our formalism in a self-consistent way. The fact of the matter is that the system of the transverse relations (3.37) has been derived.

\subsection*{Preliminary remarks}

The formalism developed in the previous subsection clearly shows that the role of the tadpole term $\Delta^2_t(D)$ in the dynamical and
gauge structures of the QCD ground state is different from those of the quark $\Delta^2_q$ and gluon $\Delta^2_g(D)$ constants.
Contrary to the gauge symmetry preservation, investigated in the first subsection of this section, the treatment
of the tadpole term cannot be put on the same footing as the quark and gluon constants, which should be omitted in the theory anyway.
Let us remind that the removal of the quark and gluon constants are due to the properties of the corresponding invariant functions, see relations (3.32),
while such an invariant function does not exist for the tadpole term itself. The derived splintering transverse relations (3.37) make it possible
for the tadpole term  $\Delta^2_t(D)$ to remain in the gluon SD eq.~(3.38) and thus it will appear in the full gluon propagator as well.
The gluon SD eq.~(3.38) is equivalent to the initial gluon SD eq.~(2.1), while eq.~(3.19) not. In other words, eq.~(3.19)
drastically distorts the true dynamical and gauge structures of eq.~(2.1), while eq.~(3.38) preserves it.
The only difference is that the sum of the skeleton loop contributions to eq.~(2.1) are taken into account in terms of the corresponding
invariant function and the tadpole term in the full gluon SD eq.~(3.38). It is much more convenient for its solution and developing
the corresponding NP MP renormalization program than the initial eq.~(2.1).
It is possible to say that the exact gauge symmetry of the QCD Lagrangian is dynamically broken in its ground state by the explicit presence
of the tadpole term in the full gluon self-energy. Precisely the system of eqs.~(3.37)-(3.39) reflects this effect.
To reveal the tadpole term's true role in full details was a clue to this discovery. Previously it has been described in some details in~\cite{8}
(and see our references therein as well).
The existence of the tadpole term is a bright signal/evidence that the real dynamical and gauge structures of the QCD ground state are not
so simple as it is required by the exact gauge symmetry of its Lagrangian.

Let us present a few important observations supporting our general statements made just above.

\begin{itemize}
\item[A.] Any deviation of the full gluon propagator from the free one requires the presence of the mass squared scale parameter
on the general dimensional ground. Even in the AF regime there is a scale violation
\begin{equation}
D_{\mu\nu}(q) \sim g_{\mu\nu} \Bigl[ { g^2 \over  1 + g^2 b_0 \ln(q^2/ \Lambda^2_{QCD}) } \Bigr] (1 / q^2),
\end{equation}
where  $g^2$ is the coupling constant and $b_0$ is the color group factor. This expression presents the summation of the so-called main PT logarithms
in QCD and written down in the 't Hooft\,--\,Feynman gauge \cite{3,4,5,8}.

\item[B.] However, the mass scale parameter $\Lambda^2_{QCD}$, which determines the non-trivial PT dynamics in the QCD vacuum, cannot be generated
by the PT itself. Due to the renormalization group equations arguments \cite{2}, any mass to which can be assigned some physical sense disappears
according to
\begin{equation}
M  \sim \mu \exp(- 1 /  b_0 g^2), \quad \quad g^2 \rightarrow 0,
\end{equation}
where $\mu$ is the arbitrary renormalization point. In other words, in every order of the PT it vanishes. So that it has to come from the IR region,
since non a finite mass can survive in the PT weak coupling limit. Such kind of mass can be of the NP dynamical origin only.

\item[C.] Due to the self-interaction of multiple massless gluon modes, QCD suffers from dangerous IR singularities (more severe than $(1/q^2)$ PT one).
The existence of the severe IR singularities also requires the presence of the mass scale parameter $\Delta^2$ in the full gluon propagator
(but "dressed" gluon remains massless), for example
\begin{equation}
D_{\mu\nu}(q) \sim T_{\mu\nu} {\Delta^2 \over (q^2)^2} \sum_{k=0}^{\infty} \Bigl({ \Delta^2 \over q^2 } \Bigr)^k \Phi_k,
\end{equation}
where the arbitrary coefficients $\Phi_k$ by themselves are the sums of the infinite number of terms~\cite{31}. This expression presents the summation
of all the possible severe IR singularities which can be taken into account by the full gluon propagator. It is worth noting that the mass scale
parameter $\Delta^2$ may contribute into the longitudinal component as well. How to deal with such severe IR singularities for the first
time has been described in~\cite{8} (and see our references therein). It will be continued in detail in the present work.

\item[D.] In general, the symmetries of the Lagrangian of the quantum/classical field theory may not coincide with the symmetries of its ground state (vacuum) and vice versa.

\item[E.] The color charge is not conserved in QCD. The derived splintering expressions (3.37) explicitly reflect this fact analytically,
while respecting the corresponding ST identities for the full gluon propagator and its free counterpart.
\end{itemize}

The invariant function $\Pi^{(s)}_t(q^2)$ associated with the transverse component of the full gluon SD eq.~(3.38), and thus with
the transverse component in the full gluon propagator, is regular at zero and only logarithmical divergent at $q^2 \rightarrow \infty$.
So it can be subject of the PT renormalization program (and it is not our problem here).
However, the tadpole term which contributes into the both components of the gluon SD eq.~(3.38), and thus contributes into the transverse and
the longitudinal component as well of the full gluon propagator, is quadratically divergent, but regularized constant. It cannot be renormalized
by the PT technics. Therefore the renormalization program beyond the PT method has to be developed (just it is our problem here). Also, in the
explicit presence of the tadpole term in the longitudinal component of the full gluon propagator the ghosts will not be able to make it
transverse since $\xi \neq \xi_0$. Therefore, the unitary of the $S$-matrix elements for the quantities calculated from first
principles will be violated, which is not acceptable. The both problems (the NP renormalization preserving the transversity) will be solved
in what follows, but before the formulation of the mass gap approach to QCD should be completed.

Concluding, a few more remarks are in order. From our analysis one can decide that not losing generality we can omit the quark degrees of freedom
below and investigate only the purely YM part of QCD. We have discussed in detail some important aspects of the color gauge structure
of the gluon SD equation in the YM gauge theory, but without any use of the PT. In obtaining these results no specific regularization schemes (preserving or not gauge invariance) has been used. No special gauge choice and no any truncations/approximations/assumptions  have been made either. Only
analytic derivations have been done, such as the decomposition into the independent tensor structures, the subtractions, etc. We have shown in the most general and unique way that the gauge symmetries of the QCD Lagrangian and its ground state are not the same, indeed.

\section{The mass gap approach to QCD}
%\label{massgap}

In order to calculate the physical observables in QCD from first principles, we need the full gluon propagator rather than the full gluon self-energy.
As emphasized earlier, the basic relations to which the full gluon propagator and its free counterpart should satisfy are the corresponding ST identities (3.1) and (3.4), respectively. They are consequence of the color gauge invariance/symmetry of QCD. However, by themselves they cannot remove the UV divergences
from the theory, as it has been shown above.
We have achieved this by formulating the suitable subtraction scheme in which the corresponding transverse conditions have been implemented.
If some equations, relations or the regularization schemes, etc.  do not satisfy them automatically, i.e., without any additional conditions, then they should be modified and not the identity (3.1). In other words, all the relations, equations, regularization schemes, etc. should be adjusted to it and not vice versa. For example, the transverse condition for the full gluon self-energy (3.37) is violated, but, nevertheless,
the general form of the ST identity (3.1) is to be maintained despite the massless or massive gluon fields are considered.
Saving the mass scale parameter in the transverse part of the full gluon propagator necessary leads to the gauge-changing
phenomenon, i.e., makes it possible to fix the function $\xi= f(\xi_0)$. It is the legitimate procedure since the full gluon propagator is
defined up to its longitudinal part only. This is true for its equation of motion as well. As explained above, the mass scale parameter is very
much needed in the transverse part of the full gluon propagator in order to correctly reflect the true dynamical structure of the QCD ground state.

Let us begin with the gluon SD eq.~(3.38), which can be equivalently re-written down as follows:

\begin{eqnarray}
D_{\mu\nu}(q) = D^0_{\mu\nu}(q) &+& D^0_{\mu\rho}(q)i T_{\rho\sigma}(q) \left[ q^2 \Pi(q^2; D) +  \Delta^2_t(D) \right] D_{\sigma\nu}(q) \nonumber\\
&+& D^0_{\mu\rho}(q)i L_{\rho\sigma}(q) \Delta^2_t(D) D_{\sigma\nu}(q),
\end{eqnarray}
where $\Pi(q^2; D)$ is a replacement (for convenience) of the initial $\Pi^{(s)}_t(q^2; D)$ in eq.~(3.38). It is worth remanding that it is regular at zero
and may have only logarithmic divergences in the PT $q^2 \rightarrow \infty$ limit.
Combining this equation with the decompositions (3.2) and (3.3), one obtains
\begin{equation}
d(q^2) = {1 \over 1 + \Pi(q^2; D) + (\Delta^2_t(D) / q^2)}.
\end{equation}
This relation is not a solution for the full gluon invariant function,
but is the NL transcendental equation for different invariant functions $d(q^2), \ \Pi(q^2; D)$ and the tadpole term $\Delta^2_t(D)$,
i.e., $d=f(D(d))$. Nevertheless, from this expression is clearly seen that in the PT $q^2 \rightarrow \infty$ regime, the mass gap term contribution
$(\Delta^2_t(D) / q^2)$ can be neglected, but the invariant function $\Pi(q^2; D)$ may still depend on $\Delta^2_t(D)$ under the PT logarithms.
At the same time, in the NP region of finite and small gluon momenta this term is dominant, and the dependence of $d(q^2)$ on $\Delta^2_t(D)$ in this case may be much more complicated due to the transcendental character of eq.~(4.2). That is why it should be kept 'alive' on a general ground, indeed.
However, keeping it 'alive' makes the YM theory unrenormalizable. So the problem arises how to make the theory renormalizable
in this case as well (see next sections).

It is instructive now to show explicitly the gauge-changing phenomenon, mentioned above. Contracting the full gluon SD eq.~(4.1)
with $q_{\mu}$ and $q_{\nu}$, and substituting its result into the general ST identity (3.1), one arrives at
\begin{equation}
q_{\mu}q_{\nu} D_{\mu\nu}(q) = i \xi_0 \left( 1 - \xi  {\Delta^2_t(D) \over q^2} \right) =  i \xi,
\end{equation}
which solution is
\begin{equation}
\xi \equiv \xi(q^2) = { \xi_0 q^2 \over q^2 + \xi_0 \Delta^2_t(D)},
\end{equation}
i.e., in this case the gauge-fixing parameter becomes the function $\xi \equiv \xi(q^2)$ and not a constant like $\xi_0$.
This is in fair agreement with our discussion just above. Let us point out that the expression (4.4) satisfies the gauge changing
condition (3.36), which is as it should be.
From eq.~(4.4) it is clearly seen that in the PT $q^2 \rightarrow \infty$ regime, which effectively equivalent to the formal $\Delta^2_t(D)=0$
limit and vice versa, the gauge-fixing parameter $\xi$ goes to $\xi_0$, as expected.
Thus, behind the general inequality $\xi \neq \xi_0$ is the tadpole term as its dynamical source, when its contribution can be neglected
only then $\xi = \xi_0$. In this connection, let us note that in the previous and this sections the equality $\xi = \xi_0$ holds
only for the regularized gluon fields. For the renormalized full gluon propagator and the free one their gauge-fixing parameters will be different
even in the formal $\Delta^2_t(D)=0$ limit. In the case of singular gluon fields investigated in this paper both types of differences
play no any role. The longitudinal component of the singular full gluon propagator will be suppressed at large distances ($q^2 \rightarrow 0$), see next
sections.

Substituting equations (4.2) and (4.4) into the general decomposition (3.2) for the full gluon propagator, one finally obtains
\begin{equation}
D_{\mu\nu}(q) =  { i T_{\mu\nu}(q) \over q^2 + q^2 \Pi(q^2; D) + \Delta^2_t(D) }  + i L_{\mu\nu}(q) { \lambda^{-1} \over q^2 + \lambda^{-1} \Delta^2_t(D)},
\end{equation}
where we have introduced the useful notation, namely $\xi_0 = \lambda^{-1}$~\cite{4} (not to be confused with the dimensionless UV regulating
parameter, mentioned in Section II). The corresponding ST identity now becomes
\begin{equation}
q_{\mu}q_{\nu} D_{\mu\nu}(q) = i \xi(q^2)  = i { \lambda^{-1} q^2 \over (q^2 + \lambda^{-1} \Delta^2_t(D))}.
\end{equation}

In what follows we call it as the generalized gauge since it depends on the tadpole term $\Delta^2_t(D)$, and when it is zero, one recovers the gauge-fixing parameter for the free gluon propagator.
To our best knowledge the generalized gauge (4.6) has been derived for the first time in a such new manner.
In the generalized gauge the gauge-fixing parameter $\lambda^{-1}$ can vary continuously from zero to infinity.
The functional dependence of the generalized gauge-fixing parameter $\xi$ (4.4) is fixed up to an arbitrary gauge-fixing parameter $\xi_0 = \lambda^{-1}$. Unless we fix it, and thus $\xi$ itself, we will call such situation as the general gauge dependence (GGD), see equations (3.1), (4.5), and (4.6).
Choosing $\xi_0 = \lambda^{-1}$ explicitly, we will call such situation as the explicit gauge dependence (EGD).
For example, $\xi_0 = \lambda^{-1}=0$ is called the unitary (Landau) gauge, $\xi_0 = \lambda^{-1}=1$ is called the t' Hooft--Feynman gauge, etc.~\cite{26,27,28,29,30}. The formal $\xi_0 = \lambda^{-1}= \infty$ limit is called as the canonical gauge in~\cite{29}, and its properties
has been discussed in detail in our preliminary work~\cite{35}. This distinction seems a mere convention, but, nevertheless, it is useful one in QCD because
of the presence of the mass scale parameter in its ground state.
The generalized gauge directly follows from the GGD/EGD formalism within the mass gap approach to QCD. It requires that there is no other functional expression for $\xi$, apart from given by the relation (4.4) at finite  $\xi_0 = \lambda^{-1}$, in the full gluon propagator (4.5)
and the corresponding ST identity (4.6) for the gluon fields in this case. In other words, unlike the expression (4.2), which is the NL transcendental relation, the expression (4.4) is the known function, which determines  $\xi = \xi(q^2)$. The system of equations, consisting of the gluon
SD eq.~(4.1), the expression (4.5) and the ST identity (4.6), constitutes that the $SU(3)$ color gauge symmetry
of the QCD Lagrangian is not a symmetry of its ground state. It is important to note that the tadpole term enters the full gluon self-energy linearly,
see, for example, the expression (2.3).
However, in the full gluon propagator (4.5) it appears in the completely different NL way. This happened because
its contribution has been summed up with the help of the gluon SD eq.~(4.1). Let us underline once more that
the tadpole term contributes into the transcendental expression, associated with the transverse component of the full gluon propagator (4.5).
At the same time, its contribution into the longitudinal component of. eq.~(4.5) renders it to the known function of $q^2$ and $\Delta^2_t(D)$.
That is why there is no equivalence between the PT $q^2 \rightarrow \infty$ and the formal $\Delta^2_t(D)=0$ limits in the former one, while this equivalence
persists in the latter one. Eq.~(4.5) shows the general structure of the full gluon propagator with the tadpole term  or without it if it is formally
put zero.

For our future aims it is convenient to present the tadpole term as follows:

\begin{equation}
\Delta^2_t(D) = \Delta^2 \times c(D), \quad \Delta^2 > 0,
\end{equation}
where $\Delta^2$ is a finite, positive mass squared scale parameter, while all the dependence in the tadpole term on the un-physical parameters of the theory,
such as the UV regulating parameter, the subtraction point, the gauge-fixing parameter, etc., are to be included into the dimensionless coefficient constant $c(D)$. It does not explicitly depend on $\Delta^2$ at any $D$ as it follows from eq.~(2.4), and see below as well.
Evidently, such factorization is always possible. However, the above-mentioned NP renormalization program will be understood as making it possible to separate $\Delta^2$ from such types of constants, which can be even more complicated than $c(D)$.
The finite constant $ \Delta^2$ will be called the mass gap, and it is nothing else but the renormalized version of the tadpole term itself.
The formal $\Delta^2_t(D)=0$ limit means that $\Delta^2=0$ and vice versa, as it follows from the relation (4.7), since the coefficient 'function'
$c(D)$ is the regularized finite quantity. The mass gap will separate the renormalized full
gluon propagator, depending on it, from its PT and free counterparts, not depending on it.

\section{Singular solution}

In the previous section it has been underlined that the relation (4.2) is not a solution for the full gluon invariant function,
but rather some kind of the NL transcendental equation for different invariant functions $d(q^2), \ \Pi(q^2; d)$ and the tadpole term $\Delta^2_t(d)$,
i.e., $d=f(d)$. It can be equivalently written down as follows:

\begin{equation}
d(q^2) = 1 - \left[ \Pi(q^2; d) + { \Delta^2_t(d) \over q^2} \right] d(q^2),
\end{equation}
indeed. Note, here and below we equivalently replaced $D$ by $d$. Its NL iteration solution has been developed in~\cite{31}.
However, here we are going to present much more economic and mathematically transparent method of its solution.
For this purpose, let us re-write the initial expression (5.1) in the slightly different but equivalent way, namely

\begin{equation}
d(q^2; z) = {1 \over 1 + \Pi(q^2; d) + z c(d)}, \quad z = { \Delta^2 \over q^2},
\end{equation}
i.e., we have already used the factorization (4.7) in order to introduce the dependence on the dimensionless variable $z$. Since the invariant function $\Pi(q^2; d)$ is regular function at small $q^2$, it can not depend on $z$, while still depending on the tadpole term or, equivalently, the mass gap itself.
Let us underline that within the mass gap approach whatever is the dependence of $\Pi(q^2; d)$ on the mass gap (even can be non-analytic due the complicate transcendental character of the relation (5.2)) the formal $\Delta^2=0$ limit for this invariant function always exists and it is $\Pi(q^2; d^{PT})$.
Then the expression (5.2) can be formally expanded in the Taylor series in powers of $z$ around zero $z$, namely

\begin{equation}
d(q^2; z) = \sum_{k=0}^{\infty} z^k f_k(q^2),
\end{equation}
where the functions $f_k(q^2)$ are the corresponding derivatives of $d(q^2; z)$ with respect to $z$ and putting after $z=0$, which implies the
formal $\Delta^2=0$ limit. This means that $d(q^2; z=0)= d^{PT}(q^2)$, while $c(d)$ is to be replaced by $c(d^{PT})$
in this formal limit. However, it remains quadratically divergent constant, but regularized one, at any $d^{PT}$. Thus, these invariant
functions are

\begin{equation}
f_k(q^2) = (-1)^k d^{PT}(q^2)[d^{PT}(q^2) c(d^{PT})]^k
\end{equation}
and

\begin{equation}
d(q^2; z=0) = d^{PT}(q^2) = f_0(q^2) = { 1  \over  1 + \Pi(q^2; d^{PT})},
\end{equation}
as it follows from the system of eqs.~(3.18)-(3.21).
It is important to emphasize that contrary to the expression (5.2), the expansion (5.3) can be considered now as a formal solution for
$d(q^2)$, since the invariant functions $f_k(q^2)$ depend on $d^{PT}(q^2)$, which is assumed to be known.

The expansion (5.3) then can be present as follows:

\begin{equation}
d(q^2; z) = \sum_{k=0}^{\infty} z^k f_k(q^2) = d^{PT}(q^2) + d^{INP}(q^2; z),
\end{equation}
where

\begin{equation}
d^{INP}(q^2; z) = \sum_{k=1}^{\infty} z^k f_k(q^2) = z \sum_{k=0}^{\infty} z^k f_{k+1}(q^2),
\end{equation}
and for the explanation of the superscript 'INP', which reminding means intrinsically NP, see next section.

The invariant functions $f_{k+1}(q^2)$ are regular functions of their argument, since they are expressed in terms of $d^{PT}(q^2)$
through the relations (5.4), and $d^{PT}(q^2)$ itself is regular function of $q^2$. Each of these functions
can be present as series in inverse powers of the variable $z$, namely $z^{-1} = (q^2 / \Delta^2)$, i.e., in powers of small $q^2$,
and the finite mass gap $\Delta^2$ determines the scale of the INP dynamics in the deep IR region. Then one obtains

\begin{equation}
f_{k+1}(q^2) = \sum_{n=0}^{\infty}  z^{-n} f^{(n)}_{k+1}(0),
\end{equation}
and substituting this series into the expansion (5.7), one arrives at

\begin{equation}
d^{INP}(q^2; z) = z \sum_{k=0}^{\infty} z^k f_{k+1}(q^2) = z \sum_{k=0}^{\infty} z^k \sum_{n=0}^{\infty}  z^{-n} f^{(n)}_{k+1}(0).
\end{equation}

After doing some algebra, this double sum series can be present as the sum of the two separate series as follows:

\begin{equation}
d^{INP}(q^2; z) = z \sum_{k=0}^{\infty} z^k \Phi_k(0) + z \sum_{k=1}^{\infty} z^{-k} \Phi_{-k}(0),
\end{equation}
where the coefficients $\Phi_k(0)$ and $\Phi_{-k}(0)$ by themselves are infinite series expressed in terms of the coefficients in the expansion (5.9).
The sum of these two series in eq.~(5.10) is nothing else but the Laurent expansion, and going now back to the variable $q^2$, one gets

\begin{equation}
d^{INP}(q^2) = \Bigl( {\Delta^2 \over q^2} \Bigr) L (q^2), %\quad q^2 \in (0, \infty),
\end{equation}
with

\begin{equation}
L(q^2) = \sum_{k= - \infty}^{\infty} \Bigl(
{\Delta^2 \over q^2} \Bigr)^k \Phi_k(0), \quad   r < {\Delta^2 \over q^2} < R,
\end{equation}
and the arbitrary finite numbers $R, r$ determine the ring of convergence of the corresponding Laurent expansion (let us remind that within the singular solution the variable $ z = \Delta^2 / q^2$ is always finite). The coefficient 'functions' (or simply the coefficients in some places above and below)
\begin{equation}
\Phi_k (0) \equiv \Phi_k(\lambda, \alpha, \xi_0, g^2) = \sum_{m=0}^{\infty} \Phi_{km}(\lambda, \alpha, \xi_0, g^2)
\end{equation}
may, in general, depend on all the possible un-physical parameters of the theory seen, for example, in the last equation
and mentioned above at the end of the previous section. Also, let us note that the sum over
index $m$ in eq.~(5.13) shows that the infinite number of the contributions invoke each term in powers of the variable $(\Delta^2 / q^2)$
in the Laurent expansion (5.12). Thus, the functional dependence of the INP invariant function $d^{INP}(q^2)$ in eq.~(5.11) is fixed up to the arbitrary
but regularized constants $\Phi_k (0)$, which appear in the corresponding Laurent expansion (5.12).

Substituting the decomposition (5.6) into the full gluon propagator (4.5), one obtains

\begin{equation}
D_{\mu\nu}(q) = D^{INP}_{\mu\nu}(q) + \tilde{D}^{PT}_{\mu\nu}(q),
\end{equation}
where

\begin{equation}
D^{INP}_{\mu\nu}(q) = i T_{\mu\nu}(q)d^{INP}(q^2) {  1 \over q^2},
\end{equation}
while $d^{INP}(q^2)$ is defined by the expressions (5.11)-(5.12), and

\begin{equation}
\tilde{D}^{PT}_{\mu\nu}(q) = i T_{\mu\nu}(q)d^{PT}(q^2) {  1 \over q^2} + i L_{\mu\nu}(q) { \lambda^{-1} \over q^2 + \lambda^{-1} \Delta^2c(d)}
\end{equation}
is the so-called spurious PT full gluon propagator,
which is legitimate to introduce, since any gluon propagator is defined up to its longitudinal component. In the PT $q^2 \rightarrow \infty$
limit it becomes

\begin{equation}
D^{PT}_{\mu\nu}(q) = i T_{\mu\nu}(q)  d^{PT}(q^2) {  1 \over q^2} + i \lambda^{-1} L_{\mu\nu}(q) { 1 \over q^2 },
\end{equation}
and thus coincides with eq.~(3.21), on account of the relation (5.5), as it needs be. Let us stress that the sum in the expression (5.14) is not a sum
of the corresponding asymtotics. The both terms in this sum are defined in the whole gluon momentum range as the full gluon propagator itself.

Let us now investigate the behavior of eq.~(5.15) for $D^{INP}_{\mu\nu}(q)$ in the limits of small and large gluon momenta. It is up to the behavior of the Laurent expansion in eq.~(5.12) beyond its ring of convergence explicitly shown in this expansion. It is well known that the Laurent expansion, being a meromorphic function, is uniformly convergent at any point of its ring of convergence \cite{32}. At the same time, in the limits $q^2 \rightarrow 0$ and $q^2 \rightarrow \infty$, i.e., beyond its ring of convergence, it has essential singularities at these limits. The behavior of any meromorphic function near its essential singularities is governed by the theorem from the theory of functions of complex variables \cite{33}, namely

\hspace{2mm}

{\bf Picard theorem:} If $z_0$ is an essential singularity of the function $f(z)$, then for any complex/real number $Z \neq \infty$, expecting, may be,
one value $Z=Z_0$, every neighborhood of $z_0$ contains infinite set of points $z$, such that

\begin{equation}
f(z) = Z, \quad z \rightarrow z_0.
\end{equation}

\hspace{2mm}

So that the Picard theorem tells us that the function $f(z)$ in the close neighborhood of its essential singularity $z_0$ can be replaced by the finite
number $Z$, which value depends only on how precisely $z \rightarrow z_0$. In order to take into account this theorem, let us
use eq.~(5.11) and re-write eq.~(5.15) in a slightly different way

\begin{equation}
D^{INP}_{\mu\nu}(q; \Delta^2) = i T_{\mu\nu}(q) {\Delta^2 \over (q^2)^2} \times L (q^2),
\end{equation}
where $L(q^2)$ is given by the expansion (5.12).
Then from the Picard theorem it follows that $L(q^2) = Z_s$ when $q^2 \rightarrow 0$ and $L(q^2) = Z_r$ when $q^2 \rightarrow \infty$, where
$Z_s$ and $Z_r$ are the corresponding numbers, which do not depend on the un-physical parameters of the theory, shown in eq.~(5.13). Their values depend
only on the way how exactly the corresponding limits are approaching. Applying these results to the INP gluon propagator (5.19), one obtains

\begin{equation}
D^{INP}_{\mu\nu}(q; \Delta^2) = i T_{\mu\nu}(q) { \Delta^2 \over (q^2)^2} \times Z_s, \quad q^2 \rightarrow 0,
\end{equation}
and

\begin{equation}
D^{INP}_{\mu\nu}(q; \Delta^2) = i T_{\mu\nu}(q) { \Delta^2 \over (q^2)^2} \times Z_r, \quad q^2 \rightarrow \infty,
\end{equation}
where subscripts "s" and "r" in $Z_s$ and $Z_r$ stand for singular and regular, respectively. In accordance with the Picard theorem, we will consider in what follows only the finite values of such type of numbers. Then the two different renormalized constants appear in the two different regimes:

\begin{equation}
\Delta^2 \times Z_s = \Delta^2_{INP}, \quad (q^2 \rightarrow 0),
\end{equation}
and

\begin{equation}
\Delta^2 \times Z_r = \Delta^2_{PT}, \quad (q^2 \rightarrow \infty),
\end{equation}
to which can be assigned the physical meaning as determining the scale of the INP and non-trivial PT dynamics in the QCD ground
state, respectively, since none of them depends on the un-physical parameters of the theory, as underlined above.
The basic element of these constants is the mass gap $\Delta^2$ itself. It is of the NP dynamical origin and comes from the IR region, as
described in Sections III and IV. The enhancement of the NP small and suppression of the PT large gluon modes shown in eqs.~(5.20)
and (5.21), respectively, with the finite corresponding constants correctly describe the true dynamical structure of the full gluon propagator.

Concluding, within the singular solution the mass gap $\Delta^2$ is always finite, but now we can consider the formal $\Delta^2=0$ limit in this solution.
From the expression (5.19), one concludes that in the formal $\Delta^2=0$ limit the Laurent expansion (5.12) has an essential singularity.
Then according to the Picard theorem in its close neighborhood, the Laurent expansion $ L(q^2) = L(q^2; \Delta^2)$ is to be replaced by the the finite number $\tilde Z$ as follows:

\begin{equation}
D^{INP}_{\mu\nu}(q; \Delta^2) = i T_{\mu\nu}(q) {\Delta^2 \over (q^2)^2} \times L (q^2) = i T_{\mu\nu}(q) {\Delta^2 \over (q^2)^2} \times \tilde Z, \quad  \Delta^2 \rightarrow 0,
\end{equation}
and therefore it goes to zero in this limit. As a result, the full gluon propagator (5.14) becomes the PT gluon propagator (5.17), as expected.
This is in complete agreement with the statement about the $\Delta^2=0$ limit made above at the beginning of this section.
This also agrees with the initial
expression (5.7) in the definition (5.17), when the variable $z = \Delta^2 / q^2$ goes to zero in this limit.

\section{INP gluon propagator}

Let us now discuss the INP gluon propagator (5.19)

\begin{equation}
D^{INP}_{\mu\nu}(q) = i T_{\mu\nu}(q) { \Delta^2 \over (q^2)^2} \times L(q^2)
\end{equation}
in more detail. Being the part of the full gluon propagator (5.14), it is defined in the whole gluon momentum range,
while its Laurent expansion

\begin{equation}
L(q^2) = \sum_{k = - \infty}^{\infty} \Bigl( {\Delta^2 \over q^2} \Bigr)^k \Phi_k(0), \quad \quad r <  {\Delta^2 \over q^2} < R
\end{equation}
is uniformly convergent only within its ring of convergence, shown above.
The most important and interesting features of the INP gluon propagator (6.1)-(6.2) are as follows:

\begin{itemize}
\item[A.] Transverse by construction, i.e., not the gauge choice by hand.
\item [B.] Its functional dependence on $(\Delta^2/q^2)$ is fixed up to the arbitrary, but regularized constants $\Phi_k(0)$.
\item[C.] The presence of the severe (or, equivalently, INP) IR singularities

$(q^2)^{-2-k}, \ k=0,1,2,3,...$ only in the $q^2 \rightarrow 0$ limit, while PT IR

singularity is $(q^2)^{-1}$, see eq.~(5.17).
\item[D.] They can only be accommodated into the transverse part (5.15) of the full gluon propagator (5.14).
They cannot be summed up to a some known function, like it has been achieved for its longitudinal component in eq.~(5.16).
\item[E.] The summation of all the possible severe IR singularities with respect to the gluon momentum.
So that all the QCD full vertices can be considered as regular functions  of all the gluon momenta involved
(see discussion at the end of Section IX as well).
\item[F.] Has an essential singularities at $ 0 \leftarrow q^2 \rightarrow \infty$, i.e., beyond its ring of convergence.
\end{itemize}

The most surprising feature of the singular solution (5.14) is that its structure at zero
($q^2 \rightarrow 0$) and at infinity ($q^2 \rightarrow \infty$), i.e., outside the ring of the uniform convergence for the Laurent expansion (6.2),
is determined by the composition $\Delta^2 /(q^2)^2$ only!, which is explicitly present in its INP part (6.1), indeed.
We will call it as the Picard effect in what follows, because of the relations (5.20)-(5.23).
Let us note that in our previous publications~\cite{34,70,71,72,73} we called this effect as the zero modes enhancement (ZME) model, since we have used it as confining ansatz for the structure of the full gluon propagator at $q^2 \rightarrow 0$. It leads to the linear rising potential between heavy quarks, see Appendix B. In other words, we have paid much more attention to the functional dependence of this composition/ansatz than to its mass scale parameter.
At that time, we have had no idea about the origin of the mass scale parameter needed for it. Its importance has been realized when we have read the JW theorem (unfortunately, rather late) after its formulation in~\cite{9}.
In this paper we have proven the existence of such mass scale parameter in the QCD ground state. It will survive the corresponding INP MP IR renormalization program and will be called the JW mass gap $\Delta^2_{JW}$. So that the ZME model becomes the mass gap approach, fully justified from the physical and mathematical points of view, see sections above and sections below, respectively.

In the $q^2 \rightarrow 0$ limit, i.e., at very large distances, the full gluon propagator (5.14) is dominated by its INP part, while all the contributions from its spurious part (5.16) are suppressed in this limit. On the contrary, in the $q^2 \rightarrow \infty$ limit, i.e., at very short distances, the full gluon propagator (5.14) will be dominated by its PT part (5.17), while its INP part will be suppressed in this limit. In other words, both regimes
in the behavior of the full gluon propagator are due to the Picard effect, indeed. At the finite gluon momentum
the full gluon propagator (4.5) may have the pole-type singularity with exactly defined gluon pole mass. This solution, called as the NP massive, is regular
at large distances ($q^2 \rightarrow 0$) and has been investigated in~\cite{35}. All this means that  the INP QCD is {\bf exactly} and {\bf uniquely} separated from the PT QCD in the full gluon propagator (5.14), despite both terms depend on the mass gap. It is important to underline that both terms in the full gluon
propagator (5.14) are not just asymptotics, since they are defined in the whole gluon momentum range $q^2 \in [0, \infty)$ as the full propagator itself.
It can not be made transverse by means of ghosts, since its longitudinal component depends on the mass gap.
At the same time, its INP part is transverse in a gauge-invariant way, as emphasized above.
Just because of all these features discussed in this section, we call this part of the full gluon propagator as the INP one. In brief, within our terminology the INP quantity, being transverse, possesses dependence on the mass gap along with the explicit presence of the severe IR singularities.

Concluding, the exact and unique separations achieved above takes place only within the single full gluon propagator. However, due to the non-abelian character of the YM theory, the different gluon propagators may interact with each other. So that in the multi-loop skeleton diagrams the mixed terms (i.e., the product
of the INP and PT parts of the different gluon propagators in the different combinations) will appear. The formulation of the INP MP IR renormalization program in Sections VIII, IX and X will explain how to deal with them. It will done in the general terms, not in detail, which in any case is beyond the scope
of the present work.

\section{The mass gap and asymptotic freedom}
%\label{euclid}
%\label{sec:6}

Let us now investigate the asymptotic properties of the full gluon propagator, defined by the system of eqs.~(5.14)-(5.17).
It is instructive to begin with its PT $q^2 \rightarrow \infty$ limit. Due to the Picard effect the INP part of the full full gluon propagator (5.14)
will be suppressed (see eq.~(5.21)) in comparison with its spurious part (5.16), which in its turn becomes the PT gluon propagator (5.17) in this limit.
So that we are left with eq.~(5.2), namely

\begin{equation}
d(q^2) = { 1 \over 1 + \Pi(q^2;d) + (\Delta^2_t(d) / q^2)},
\end{equation}
and it is worth reminding that $\Delta^2_t(d) = \Delta^2 c(d)$ is the tadpole term, which is the quadratically divergent
mass scale parameter, but regularized one.
The invariant function $\Pi(q^2; d)$, being regular function at small $q^2$, may have the logarithmic divergences only in the PT $q^2 \rightarrow \infty$ limit.
Thus, one can neglect the $(\Delta^2_t(d) / q^2)$ contribution in this limit in the previous eq.~(7.1), indeed.
Let us also put for further convenience $d(q^2) =  \alpha_s(q^2; \Lambda^2) / \alpha_s(\lambda)$, where $\Lambda^2$ and $\lambda$ are the UV regulating parameter and its dimensionless counterpart, respectively, mentioned in Section II, while both $\alpha_s(q^2; \Lambda^2)$ and $\alpha_s(\lambda)$ present the PT running effective charges. Then for the PT running effective charges from eq.~(7.1), one obtains

\begin{equation}
\alpha_s(q^2; \Lambda^2) = { \alpha_s(\lambda) \over 1 + b_0 \alpha_s(\lambda) \ln (q^2 / \Lambda^2)},
\end{equation}
where, as usual, we factorized the dependence on $b_0$ and $\alpha_s(\lambda)$ in the logarithmical divergent invariant function to leading order.
This is nothing else but the sum of the main PT logarithms~\cite{3,4,5,8,39}. Here $b_0= 11/4 \pi > 0$ is the color group factor for the YM fields.
Evidently, in this regularized expression one can not go directly to the $(\Lambda^2, \lambda \rightarrow \infty)$ limits in order to finally arrive at the renormalized expression. Some intermediate steps are to be performed in order to complete the corresponding renormalization program.

Due to the Picard effect discussed in Section V, we already know that there exists the finite mass squared parameter defined in eq.~(5.23), namely $\Delta^2_{PT} = \Delta^2 \times Z_r$,
which survives the PT limit, see eq.~(5.21). It is basically determined by the mass gap $\Delta^2$ itself, which dominates the dynamical structure
of the full gluon propagator in the IR region, see eq.~(5.2).
This effect explicitly shows that in some cases it is useful to distinguish between the suppression of the
contribution $(\Delta^2_t(d) / q^2)$ in the PT limit at finite $\Delta^2_{PT}$, which appear in this limit, i.e., the term itself can be
suppressed/neglected but the constant itself exists and becomes finite.
It is possible to say that the Picard number $Z_r$ (5.23) pulls out the mass gap $\Delta^2$ from the NP region of small gluon momenta to the PT region of large gluon momenta. All this means that within our approach the UV regulating parameter $\Lambda^2$ can be always replaced by

\begin{equation}
\Lambda^2 = f(\lambda) \Delta^2_{PT},
\end{equation}
where $f(\lambda)$ is the corresponding dimensionless function, which is quadratically divergent in the $\lambda \rightarrow \infty$ limit, but otherwise
remains arbitrary. Let us emphasize that such kind of the relation makes sense to use only because we know in advance that the finite constant $\Delta^2_{PT}$ exists and comes out into the play in the PT limit, see eq.~(5.23).
Substituting this relation into eq.~(7.2) and doing some algebra, one obtains

\begin{equation}
\alpha_s(q^2) = { \alpha_s \over 1 + b_0 \alpha_s \ln (q^2 / \Delta^2_{PT})},
\end{equation}
if and only if

\begin{equation}
\alpha_s = { \alpha_s(\lambda) \over 1 - b_0 \alpha_s(\lambda) \ln f(\lambda)}
\end{equation}
exists and is finite in the weak coupling limit $\alpha_s(\lambda) \rightarrow 0$ when $\lambda \rightarrow \infty$.
The finite $\alpha_s$ can be identified with the fine structure constant of the strong interactions, calculated at some fixed scale,
$\alpha_s \equiv \alpha_s(M_Z) = 0.1184$~\cite{40}.
The existence and finiteness of $\alpha_s$ can be shown within the standard renormalization group equations arguments. Indeed, from eq.~(7.5)
it follows

\begin{equation}
\ln f(\lambda) = { \alpha_s - \alpha_s(\lambda) \over  \alpha_s b_0 \alpha_s(\lambda)} \rightarrow { 1 \over b_0 \alpha_s(\lambda)},
\quad \alpha (\lambda) \rightarrow 0, \quad \lambda \rightarrow \infty,
\end{equation}
which means that

\begin{equation}
f(\lambda) = \exp ( 1 / b_0 \alpha_s(\lambda))
\end{equation}
in the weak coupling limit shown in eq.~(7.6). Substituting this into the relation (7.3), it becomes

\begin{equation}
\lim_{(\Lambda^2, \lambda) \rightarrow \infty} \Lambda^2 \exp \left[ - { 1 \over b_0 \alpha_s(\lambda)} \right] = \Delta^2_{PT},
\quad \alpha(\lambda) \rightarrow 0,
\end{equation}
which is the finite limit of the renormalization group equations solution~\cite{39}. Note that with $\lambda \rightarrow \infty$ the effective charge
$\alpha_s(\lambda)$ has to have the logarithmical fall off (weak coupling limit) in order to satisfy to the final limit in eq.~(7.8).
It is easy to check that such behavior justifies the existence of the finite $\alpha_s$ in eq.~(7.5) as well.

At very large gluon momentum $q^2$ from eq.~(7.4) one recovers

\begin{equation}
\alpha_s(q^2) = { 1 \over b_0 \ln (q^2 / \Lambda^2_{YM})} \rightarrow 0, \quad q^2 \rightarrow \infty,
\end{equation}
which is AF famous formula~\cite{3,4,5,8,39}. In order to get this expression from eq.~(7.4) we identify $\Delta^2_{PT}$ there with $\Lambda^2_{YM}$, which is $\Lambda^2_{QCD}$ for the purely YM fields. Evidently, this can be always done by redefining the dimensionless function $f(\lambda)$ in eq.~(7.3), since
$\Lambda^2 = f(\lambda) \Delta^2_{PT}= f(\lambda) A \Lambda^2_{YM}= \tilde{f}(\lambda) \Lambda^2_{YM}$, and $A$ is the arbitrary finite number.
In fact, at $q^2$ sufficiently large $\Delta^2_{PT}$ can be replaced by $\Lambda^2_{YM}$, or simply these two constants can be identified, indeed.

The AF expression (7.9) shows that in the AF regime the every excitation of vacuum has the energy squared $q^2 > \Delta^2_{PT}$.
The explicit presence of the mass gap in the full gluon propagator
makes it possible to explain the scale violation in the PT regime. The mass gap coming from the IR region, nevertheless, survives the PT
renormalization program and finally becomes $\Lambda^2_{YM}$ in the weak coupling limit as described above and compare with eq.~(3.40).
$\Lambda^2_{YM}$ determines the scale of the non-trivial PT dynamical structure of the QCD/YM ground state. Let us remind that the PT can not
generate a mass, see eq.~(3.41), i.e., the question where the scale $\Lambda^2_{YM}$ comes from? is not answered by the PT itself.
The mass gap approach solves this problem, namely the mystery of the scale violation (which is completely NP effect) in the AF regime,
without introducing some extra degrees of freedom into the theory.

Concluding, all this is a manifestation that "the problems encountered in perturbation theory are not mere mathematical artifacts but rather signify
deep properties of the full theory"~\cite{41}. The message that we are trying to convey is that the INP dynamical structure of the full gluon propagator
indicates the existence of its non-trivial perturbative  structure and vice versa. Both effects are due to the explicit presence of the mass gap in the
dynamical structure of the QCD ground state, and are subjects of its two different renormalization programs. Let us also note in advance
that both effects AF and gluon confinement (see Appendixes B and C) require the corresponding 'running' effective charges.
However, the effective charges becomes 'running' if and only if the mass squared scale parameter exists in the ground state.

\section{Theory of distributions and dimensional regularization}

In order to investigate the IR asymptotic ($q^2 \rightarrow 0$) limit of the full gluon propagator (5.14), it is convenient to begin with some
general discussion of the properties of the gauge particles Green's functions in the IR region.
In principle, all the Green's functions (propagators) in QCD are generalized
functions, i.e., they are distributions~\cite{36}. This is especially true
for the severe IR singularities of the full gluon propagator due to
the self-interaction of massless gluons in the QCD vacuum, discussed and derived above.
They present a rather broad and important class of functions with
algebraic singularities, i.e., functions with non-summable
singularities at isolated points (at zero in what follows).
Roughly speaking, this means that all the relations involving
distributions should be considered under corresponding integrals,
taking into account the smoothness properties of the corresponding
space of test functions. Let us note in advance that the
space in which our generalized functions are continuous linear
functionals is $K$, that's the space of infinitely differentiable functions having compact support,
i.e., they are zero outside some finite region (different for each differentiable function).

The INP gluon propagator (6.1), being the part of the full gluon propagator, is defined in the whole gluon momentum range.
At the same time, the corresponding Laurent expansion (6.2) is uniformly convergent only at any point of its ring of convergence.
Thus, the INP gluon propagator (6.1) contains the infinite number of severe IR singularities $\sim (q^2)^{-2-k}, \ k=0,1,2,3,...$
when $q^2 \rightarrow 0$, i.e., beyond its ring of convergence. We will show below that they cannot be treated as the standard PT IR singularity,
which is as much singular as $\sim (q^2)^{-1}$ only. The theorem, which allows to put the severe IR singularities under control, is as follows~\cite{36}:

\hspace{2mm}

{\bf Gelfand-Shilov (GS) theorem:} For the positive quadratic form

\begin{equation}
P(q) = q^2_0 + q^2_1 +q^2_2 +...+ q^2_{n-1} = q^2 > 0,
\end{equation}
and $n$ is even number, let us define the generalized function $P^{\lambda}$ as follows:

\begin{equation}
(P^{\lambda}, \varphi) = \int d^dq P^{\lambda}(q) \varphi(q),
\end{equation}
where $\lambda$ is a complex number and $d$ is the dimension of the loop integral. The function
$\varphi(q)$ is the above-mentioned some test function. At $Re \lambda \geq 0$ this integral
is convergent and analytic function of $\lambda$. For $Re \lambda < 0$  this integral has a simple poles at points

\begin{equation}
\lambda = - {n \over 2} - k, \quad k=0,1,2,3...
\end{equation}
so that the distribution $P^{\lambda}(q)$ itself can be present as

\begin{equation}
P^{\lambda}(q) = (q^2)^{\lambda} = { C^{(k)}_{-1} \over \lambda + (d/2) +k } + finite \ terms,
\end{equation}
where residues are

\begin{equation}
 C^{(k)}_{-1} = { \pi^{n/2} \over 2^{2k} k! \Gamma((n/2) + k) } \cdot D^k \delta^n (q),
\end{equation}
while $D$ denotes d'Alembert/Laplace operator in Euclidean metric, so it is

\begin{equation}
D = { \partial^2 \over \partial q_0^2} + { \partial^2 \over \partial q_1^2} +...+ { \partial^2 \over \partial q^2_{n-1} }.
\end{equation}

\hspace{4mm}

In the dimensional regularization method (DRM)~\cite{26} the UV divergences and IR singularities are regulated by admitting that
$d = n - 2 \delta, \ \delta \rightarrow 0^+$ and

\begin{equation}
d = n + 2 \epsilon, \quad \epsilon \rightarrow 0^+,
\end{equation}
respectively. Then from eq.~(8.4), on account of equations (8.3) and (8.7), it follows that

\begin{equation}
\lambda + {d \over 2} + k = \epsilon, \quad \epsilon \rightarrow 0^+,
\end{equation}
and the corresponding expression (8.4) can be now treated as the dimensional regularization expansion (DRE) in powers of $\epsilon$, namely

\begin{equation}
(q^2)^{\lambda} = { 1 \over \epsilon } C^{(k)}_{-1}  + finite \ terms , \quad \lambda = - (n / 2) - k, \ k=0,1,2,3... , \quad \epsilon \rightarrow 0^+.
\end{equation}
%\begin{equation}
%(q^2)^{-˙{n \over 2} - k} = { 1 \over \epsilon } C^{(k)}_{-1}  + finite \ terms, \quad \epsilon \rightarrow 0^+.
%\end{equation}
where we can put $d=n$ now (i.e., after introducing this
expansion). By the "$finite \ terms$" here and everywhere a number of
necessary subtractions under corresponding integrals is understood
\cite{36} (see below as well). Let us underline its most remarkable feature. The order of
singularity does not depend on $n$ and $k$ (and thus on $\lambda$). In terms of
the IR regularization parameter $\epsilon$ it is always a simple
pole $1/ \epsilon$. However, the residue at a pole will be drastically
changed from one power singularity to another. This means
different solutions to the whole system of the dynamical equations for
different set of numbers $\lambda$ and $k$. Different solutions
mean, in their turn, different vacua. In this picture different
vacua are to be labeled by the two independent numbers: the
exponent $\lambda$ and $k$. At a given number of $d(=n)$ the
exponent $\lambda$ is always negative being integer if $d(=n)$ is
an even number or fractional if $d(=n)$ is an odd number. The
number $k$ is always integer and positive and precisely it
determines the corresponding residue at a simple pole, see
eq.~(8.5). It would not be surprising if these numbers were somehow
related to the nontrivial topology of the QCD vacuum in any dimensions, though
by themselves they characterize its dynamical structure.

It is worth emphasizing that the structure of the severe IR
singularities in Euclidean space is much simpler than in Minkowski
space, where kinematical (un-physical) singularities due to the
light cone also exist \cite{2,36} (in this connection let us
remind that in Euclidean metric $q^2 = 0$ implies $q_i=0$ and
vice-versa, while in Minkowski metric this is not so). In this
case it is rather difficult to untangle them correctly from the
dynamical singularities, the only ones which are important for the
calculation of any physical observable from first principles. The consideration is
much more complicated in the configuration space \cite{36}. That is
why we prefer to work in the momentum space (where
propagators do not depend explicitly on the number of dimensions)
with Euclidean signature. We also prefer to work in the covariant
gauges in order to avoid peculiarities of the non-covariant gauges
\cite{2,37,38}, for example, how to untangle the gauge pole from
the dynamical one.

In principle, none of the regularization schemes (how to introduce
the IR regularization parameter in order to parameterize severe IR
singularities and thus to put them under control) should be
introduced by hand. Some of them have been cited in appendix 3.B of our book \cite{8}.
However, not the regularization only is important but also its combination with the distribution theory.
Just this provides an adequate mathematical
framework for the correct treatment of all the Green's functions
in QCD. The regularization of the INP IR singularities in four-dimensional QCD is determined
by the Laurent expansion (8.9) at $n=4$ as follows:

\begin{equation}
(q^2)^{- 2 - k } = { 1 \over \epsilon} a(k)[\delta^4(q)]^{(k)} +
f.t. = { 1 \over \epsilon} \Bigr[ a(k)[\delta^4(q)]^{(k)} +
O_k(\epsilon) \Bigl], \quad \epsilon \rightarrow 0^+,
\end{equation}
where $a(k)= \pi^2 / 2^{2k} k! \Gamma(2+k)$ is a finite constant depending only on $k$ and $[\delta^4(q)]^{(k)}$ represents
the $k$'s derivative of the $\delta$-function, see eqs.~(8.5)-(8.6). We point out once more that after introducing this expansion
everywhere one can fix the number of dimensions, i.e., put $d=n=4$ for QCD without any further problems. Indeed there will be no
other severe IR singularities with respect to $\epsilon$ as it goes to zero, but those explicitly shown in this expansion.
In this connection, let us remind that the dependence on the initial $\alpha$ as well as on other un-physical parameters disappears due the Picard theorem.
The finite terms in the DRE (8.10) is nothing else but its regular part, and it is

\begin{equation}
f.t. =(q^2)^{- 2 - k }_{-} + \epsilon (q^2)^{- 2 - k }_{-} \ln q^2 + O(\epsilon^2),
\end{equation}
where the functional $(q^2)^{- 2 - k }_{-}$ determines the number of the above-mentioned subtractions, see~\cite{36}. These
terms, however, do not undermine the INP MP IR renormalization program which will be
discussed below. The DRE (8.10) takes place only in four-dimensional QCD with
Euclidean signature. In other dimensions and/or Minkowski
signature it is much more complicated as pointed out above. As it
follows from this expansion any power-type INP IR singularity,
including the simplest one at $k=0$, scales as $1 /\epsilon$ as it
goes to zero. Just this plays a crucial role in the INP IR
renormalization of the theory within our approach. Evidently, such
kind of the DRE (8.10) does not exist for the PT IR singularity $(q^2)^{-1}$. This is one more important criterion
which makes the separation between the INP and PT parts of the full gluon propagator (5.14), discussed above, unique and exact.

As pointed out in \cite{4}, the general problem of the loop integrals such as

\begin{equation}
\int { d^d q \over (2 \pi)^d} {q_{\mu_1}...q_{\mu_p} \over (q^2)^m},
\end{equation}
which appear in QCD calculations is that they are ill-defined. There is no dimension where they are meaningful.
They are either IR singular if $d+p-2m \leq 0$ or UV divergent if $d+p-2m \geq 0$, i.e., depending
on the relation between the numbers $d$, $p$ and $m$. However, from our presentation above it follows that such kind of the IR singularities
can be now properly dealt with by the joint use of the distribution theory and the DRM. As we already know, such
tadpole-type loop integrals can be disregarded in the PT limit (see discussion in~\cite{4,8} as well) or they are simply convergent (according to the GS theorem).

In summary, first we have emphasized the
distribution nature of the INP (severe) IR singularities. Secondly, we have
explicitly shown how the DRM should be correctly and in a self-consistent way implemented into the distribution theory.
This makes it possible to put the severe IR singularities under a firm mathematical control.

\section{INP MP IR renormalization program}

Before starting to formulate the INP MP IR renormalization program to make theory finite when the IR dimensional regularization
parameter $\epsilon$ goes to zero
at final stage, let us remind the two important results obtained in our investigation. First, in the DRE (8.10) the pole over $\epsilon$ is always a simple pole  $1 / \epsilon$, and does depend on the number $k$ in its right-hand-side. It is easy to understand that just this makes the theory IR renormalizable; otherwise it would be unrenormalizable. The second point is that due to the Picard theorem only the simplest INP IR singularity at $k=0$ survives in the theory
when $q^2 \rightarrow 0$. Keeping in mind these two important observations, it is instructive to present the full gluon propagator
(5.14)-(5.16), on account of eq.~(6.1), as follows:

\begin{equation}
D_{\mu\nu}(q) =   {  i \over (q^2)^2} \Bigl[ T_{\mu\nu}(q) \Delta^2 L(q^2) + q^2 R_{\mu\nu}(q) \Bigr], \quad q^2 \in [0, \infty),
\end{equation}
where

\begin{equation}
R_{\mu\nu}(q) = T_{\mu\nu} d^{PT}(q^2) +  L_{\mu\nu}(q) { q^2 \lambda^{-1} \over q^2 + \lambda^{-1} \Delta^2c(d)},
\end{equation}
and thus $R_{\mu\nu}(q)$ is a regular function of $q^2$ when it goes to zero, while the purely longitudinal component of the full gluon propagator
(9.1) simply disappears in this limit as it follows from the definition (9.2). In other words, the full gluon propagator (9.1) becomes purely
transverse in the $q^2 \rightarrow 0$ limit, i.e., at large distances.

The next step is to explicitly write down the DRE (8.10) for the simplest NP IR singularity at $k=0$

\begin{equation}
(q^2)^{- 2} = { 1 \over \epsilon} a(0)[\delta^4(q)] + f.t. = { 1 \over \epsilon} \Bigr[ a(0) \delta^4(q) +
O(\epsilon) \Bigl], \quad \epsilon \rightarrow 0^+,
\end{equation}
where $a(0) = \pi^2$. Due to the $\delta^4(q)$-function in the residue of this DRE, all the functions which appear under corresponding
skeleton loop integrals should finally be replaced by their values at $q=0$. Substituting this expansion into the full gluon
propagator (9.1), and after doing some algebra, one obtains

\begin{equation}
D_{\mu\nu}(q) = {1 \over \epsilon} \Bigl[ i T_{\mu\nu}(q) \pi^2 \Delta^2 Z_s \delta^4(q)+ O_{\mu\nu}(q; \epsilon) \Bigr], \quad \epsilon \rightarrow 0^+,
\end{equation}
which is nothing else but the DRE for the full gluon propagator (9.1), and we replace $L(q^2)$ by the constant $Z_s$ in accordance with the relations (5.19)-(5.21). Here $O_{\mu\nu}(q; \epsilon)$ denotes the tensor terms of the $\epsilon$-order.

As it follows from the general DRE (8.10), whatever the severe IR singularity is in terms of $q^2$, in terms of the IR
regularization parameter $\epsilon$ it always scales as  $1 / \epsilon$. The IR renormalized full gluon propagator $D^R_{\mu\nu}(q)$,
which should exist or vanish as $\epsilon \rightarrow 0^+$, is to be defined as follows:

\begin{equation}
D^R_{\mu\nu}(q) = X(\epsilon) D_{\mu\nu}(q),
\end{equation}
and $X(\epsilon)$ is the INP MP IR renormalization constant for the full gluon propagator or, equivalently, the gluon wave function INP MP IR
renormalization one. Its general expansion in terms of the positive powers in $\epsilon$ has to be

\begin{equation}
 X(\epsilon) = A \epsilon + O(\epsilon^2), \quad \epsilon \rightarrow 0^+,
\end{equation}
where $A$ and similar quantities below can be any complex/real numbers. Evidently, such expansion for the INP MP gluon wave function IR renormalization constant makes the full gluon propagator (9.4) finite in the final $\epsilon \rightarrow 0^+$ limit. Thus, for the IR renormalized full gluon propagator (9.5),
one finally arrives at ($\epsilon \rightarrow 0^+$)

\begin{equation}
D^R_{\mu\nu}(q) = i T_{\mu\nu}(q) \pi^2 \Delta^2_R \delta^4(q)+ \tilde{O}_{\mu\nu}(q; \epsilon), \ \textrm{ $q^2$
is an independent loop variable},
\end{equation}
on account of eq.~(9.4), and where we introduce the following notation $\Delta^2_R = \Delta^2 Z_s A$, for convenience. Here $\tilde{O}_{\mu\nu}(q; \epsilon)$
denotes the new tensor terms of the $\epsilon$-order, while the tensor terms of the $\epsilon^2$-order are omitted, for convenience.

However, this is not the whole story yet. In the multi-loop skeleton diagrams the more severe IR singularities than the simplest
INP IR singularity may appear. Then the product of at least two $\delta$-functions at the same point becomes possible, if directly
substituting eq.~(9.7) for the different full gluon propagators. Such kind of the product is ill-defined in the theory of
distributions \cite{36}. In order to avoid this problem, the IR renormalized full gluon propagator (9.5) should be used in its general form as follows:

\begin{equation}
D^R_{\mu\nu}(q) = X(\epsilon) \times {  i \over (q^2)^2} \Bigl[ T_{\mu\nu}(q) \Delta^2 L(q^2) + q^2 R_{\mu\nu}(q) \Bigr],  \quad \epsilon \rightarrow 0^+.
\end{equation}
It is easy to see that if $q^2$ is independent loop variable then applying the DRE (8.10) for $k=0$, i.e., eq.~(9.3), one again arrives
at the expression (9.7), on account of the expansion (9.6).
Using this general expression for the IR renormalized full gluon propagator in the multi-loop skeleton diagrams instead of the $\delta$-functions
in the residues, their derivatives will appear according to the DRE (8.10). How to treat the loop integrals with the derivatives of the $\delta$-functions
is well-known procedure in the theory of the generalized functions (distributions)~\cite{36} (these integrals should be complemented by the corresponding
number of the subtractions at non-zero number $k$). The careful separation of the severely singular terms from the regular ones in the multi-loop skeleton diagrams is also important (in a single full gluon propagator this is automatically achieved, see eq.~(9.8)). However, as emphasized above,
the IR renormalization of the theory will not be undermined
in this case as well, since a pole in $\epsilon$ is always a simple pole $1 / \epsilon$ whatever INP IR singularity with respect to $q^2$ appears in the theory, see again DRE (8.10).
The expression (9.8) for the IR renormalized full gluon propagator does not depend whether the gluon
momentum is the independent skeleton loop variable or not. Indeed, if it approaches zero as a loop variable then the expression (9.7) is to be
used, as described above. At the same time, if the gluon momentum approaches zero as a free particle variable, i.e., it is an external (not a loop) variable, then the expression (9.8) cannot be treated as the distribution. So that it is a standard function and the DRE (8.10) is not the case to be used,
and thus eq.~(9.8) disappears as $\epsilon$ in the $\epsilon \rightarrow 0^+$ limit, again on account of the expansion (9.6), namely

\begin{equation}
D^R_{\mu\nu}(q) = X(\epsilon) \times  {  i \over (q^2)^2} \Bigl[ T_{\mu\nu}(q) \Delta^2 L(q^2) + q^2 R_{\mu\nu}(q) \Bigr]  \sim \epsilon, \ \textrm{ $q^2$ is not a loop variable},
\end{equation}
i.e., it is a free particle variable. In the $q^2 \rightarrow 0$ limit, we again can replace $L(q^2) \rightarrow Z_s$ and suppress $q^2 R_{\mu\nu}(q)$ term.
This behavior is gauge-invariant, does not depend on any truncations/approximations/assumptions, based on a firm mathematical grounds, and thus it is a general one. It prevents the transverse full ("dressed") gluons from appear in asymptotic states (large distances) as physical
particles. So color gluons can never be isolated, remanding that NP massive~\cite{35} and free gluons are suppressed at large distances ($q^2 \rightarrow 0$) and may exist at short and very short distances only. This solves a long-standing color gluon confinement problem within the mass gap approach to QCD.
As pointed out in our book~\cite{8} (see discussion and references therein), we came back to the old IR slavery mechanism of gluon confinement but on a new basis ('new is well-forgotten old'). We put this mechanism on a firm mathematical ground in order to control severe IR singularities in QCD due to the
self-interaction of the massless gluon modes. In the past the IR slavery mechanism has been abandoned just because nobody knew how to deal with
such severe IR singularities in the correct mathematical way.

Moreover, our findings here
make the resulting  YM theory effectively abelian. Firstly, all the severe IR singularities with respect to the gluon momenta involved have been summarised
into the full gluon propagator, so that all the QCD vertices can be considered as regular functions of their gluon variables.
If not, which is very unlikely in our opinion, since they have been summed up to infinity, then, nevertheless, the multiplication of the Laurent series, depending on the gluon momenta coming from the different sources (the gluon propagators themselves and the corresponding vertices), is only a technical
mathematical problem. After the multiplication is performed, the separation of the severely singular terms from the regular ones should be proceed in the same way, as described above. So that the vertices effectively become the regular quantities of the gluon momenta involved (the multiplied
Laurent series is a Laurent series again, and all the dependence on the gluon momenta can be effectively attributed to the Laurent expansions of the corresponding full gluon propagators, indeed). However, the most important second observation, which justifies the first one as well, is as follows: due
to the behavior of eq.~(9.9) only those multi-loop skeleton contributions in the resulting theory will survive where the number of the independent gluon skeleton loop variables is equal to the number of the full gluon propagators involved. In all other cases their contributions will be suppressed at least as $\epsilon$ when it goes to zero at final stage. So the resulting QCD effectively becomes abelian-type theory. It is possible to say that all the non-abelian degrees of freedom have been incorporated into the mass gap, via the composition $\Delta^2 / (q^2)^2$, and the invariant function $L(q^2)$ in the transverse
full gluon propagator (9.8).

Concluding, it is instructive to note that the final different results, reproduced in equations (9.7) and (9.9) for the IR renormalized
full gluon propagator (9.8), depend on the gluon momentum whether it approaches zero as a loop variable or as a free particle
variable, respectively. This is in complete agreement with the Picard theorem, which separates these two principally different cases, indeed.
The full gluon propagator (9.1), which contain the INP part (6.1), is also can be called the INP singular full gluon propagator.
Its IR renormalized version is eq.~(9.8), and it is defined in the whole gluon momentum range $q^2 \in [0, \infty)$, i.e., it is not just asymptotic at
$q^2 \rightarrow 0$.

\hspace{2mm}

\section{IR renormalizations of the ST identity and the tadpole term}

\hspace{2mm}

In order to go further and complete the renormalization program it is necessary to consider the important issue of the
INP MP IR renormalization of the ST identity. It determines the longitudinal component of the full gluon propagator. The IR renormalizazion program
for its transverse component has been performed just in the previous section.
Though both components are independent from each other, but, nevertheless,
their renormalization programs have to be agreed in any case.
Otherwise the general IR renormalizability of the theory will be destroyed. From eqs.~(9.1)-(9.2), one obtains

\begin{equation}
q_{\mu}q_{\nu} D_{\mu\nu}(q) = i { \lambda^{-1} q^2 \over q^2 + \lambda^{-1} \Delta^2_t(d)},
\end{equation}
which coincides with the longitudinal component of the singular solution (9.1)-(9.2), as it needs be.
Let us note that in eqs.~(9.1)-(9.2) it is suppressed in comparison with its transverse counterpart in the $q^2 \rightarrow 0$ limit. However, it
determines the structure of the ST identity, as it follows from the previous relation. That is why it is important to fix the relation between
the INP IR renormalization constants for the gluon propagator and the tadpole term itself. So that, introducing such IR renormalization constant
$X'(\epsilon)$ for the tadpole term as follows:

\begin{equation}
\Delta^2_t(d)  =  X'(\epsilon) \bar \Delta^2,  \quad \epsilon \rightarrow 0^+,
\end{equation}
and doing some algebra, one finally obtains

\begin{equation}
q_{\mu}q_{\nu} \bar D_{\mu\nu}(q) = i {\bar \lambda^{-1} q^2 \over q^2 + \bar \lambda^{-1} \bar \Delta^2},
\end{equation}
where  $\bar \lambda^{-1} = X'(\epsilon) \lambda^{-1}$, while

\begin{equation}
\bar D_{\mu\nu}(q) = X'(\epsilon) D_{\mu\nu}(q),
\end{equation}
and here and below all the quantities with bar exist as $\epsilon \rightarrow 0^+$.

From the previous section, we already know that the full gluon propagator $D_{\mu\nu}(q)$ is always scales as $1/\epsilon$.
Then in complete agreement with the definition (9.5) and the expansion (9.6), the expansion for $X'(\epsilon)$ in the definition (10.4) is to be

\begin{equation}
 X'(\epsilon) = A' \epsilon + O(\epsilon^2), \quad \epsilon \rightarrow 0^+.
\end{equation}

In order to proceed further, let us investigate the IR renormalization of the full gluon propagator (9.1) in more detail, which is

\begin{equation}
D_{\mu\nu}(q) =   {  i \over (q^2)^2} \Bigl[ T_{\mu\nu}(q) \Delta^2 L(q^2) + q^2 R_{\mu\nu}(q) \Bigr], \quad q^2 \in [0, \infty),
\end{equation}
where $R_{\mu\nu}(q)$ is defined in eq.~(9.2). Recollecting that the explicit expression for the Laurent expansion $L(q^2)$ is

\begin{eqnarray}
L(q^2) = \sum_{k = - \infty}^{\infty} \Bigl( {\Delta^2 \over q^2} \Bigr)^k \Phi_k(0) &=& \sum_{k = 0}^{\infty} \Bigl( {\Delta^2 \over q^2} \Bigr)^k \Phi_k(0) + \sum_{k = 1}^{\infty} \Bigl( {q^2 \over \Delta^2} \Bigr)^k \Phi_{-k}(0) \nonumber\\
&=& L_s(q^2) + L_r(q^2),
\end{eqnarray}
and we present the initial Laurent expansion as a sum of its singular (subscript 's') and regular (subscript 'r') parts when $q^2 \rightarrow 0$.
The regular part can be further present as follows:

\begin{equation}
L_r(q^2) = \sum_{k = 1}^{\infty} \Bigl( {q^2 \over \Delta^2} \Bigr)^k \Phi_{-k}(0) = \Bigl( {q^2 \over \Delta^2} \Bigr) \sum_{k = 0}^{\infty}
\Bigl( {q^2 \over \Delta^2} \Bigr)^k \Phi_{-k-1}(0)= \Bigl( {q^2 \over \Delta^2} \Bigr)f(q^2),
\end{equation}
while $f(q^2)$, in general, is a some regular function of its argument when it goes to zero.

Substituting further the decomposition (10.7), on account of eq.~(10.8), into eq.~(10.6) and doing some algebra, one obtains

\begin{equation}
D_{\mu\nu}(q) =   {  i \over (q^2)^2} \Bigl[ T_{\mu\nu}(q) \Delta^2 L_s(q^2) + q^2 \tilde{R}_{\mu\nu}(q) \Bigr],
\end{equation}
where

\begin{equation}
\tilde{R}_{\mu\nu}(q) = T_{\mu\nu} [d^{PT}(q^2) + f(q^2) ] +  L_{\mu\nu}(q) { q^2 \lambda^{-1} \over q^2 + \lambda^{-1} \Delta^2c(d)}.
\end{equation}
The function  $f(q^2) \sim \Delta^2 / q^2$ in the
PT $q^2 \rightarrow \infty$ limit as it comes out from eq.~(5.21). So that in this limit the last expression
goes to the correct PT expression. The expression (10.4) for the full gluon propagator clearly
shows that we have achieved the exact and unique separation between its INP singular and regular parts in the $q^2 \rightarrow 0$ limit.
Neglecting the regular part $\tilde{R}_{\mu\nu}(q)$  in this limit, then from eqs.~(10.9) and (10.10) one arrives at

\begin{equation}
D_{\mu\nu}(q) =  { \Delta^2 \over (q^2)^2} i T_{\mu\nu}(q) \sum_{k = 0}^{\infty} \Bigl( {\Delta^2 \over q^2} \Bigr)^k \Phi_k(0),
\quad q^2 \rightarrow 0.
\end{equation}
Let us remind that all the severe IR singularities $(q^2)^{-2-k}, \ k=0,1,2,3...$, presented in eq.~(10.11), scale as $1 / \epsilon$ in the final $\epsilon \rightarrow 0^+$ limit. In principle, all the remaining parameters of the theory (the mass gap $\Delta^2$ and the coefficients $\Phi_k(0)$) may depend on $\epsilon$. However, these coefficient 'functions' do not depend on the mass gap $\Delta^2$. Secondly, they are of the PT origin, and, therefore, they cannot
singularly depend on $\epsilon$, which may only appear in the case of the severe IR singularities, see the expansion (8.10). It is worth to explain this
point in a bit more detail. The above-mentioned coefficient functions $\Phi_k(0)$ are the multi-loop integrals which are symbolically present in the expression (8.12). So that the combination $d+p-2m \geq 0$ is always valid, since the PT gluon propagators behave like $1/q^2$ in the $q^2 \rightarrow 0$ limit.
Thus the coefficient functions $\Phi_k(0)$ are free from the severe IR singularities, as underlined above, while remaining the finite quantities at zero. Therefore not losing generality, one can consider them as not depending on $\epsilon$ at all. We are left with the mass gap $\Delta^2$ only,
but due to the decomposition (4.7) it has been defined as the finite quantity, not depending on anything.

However, it is worth remembering now that the expansion (10.11) has been obtained by the differentiation with respect to variable $z=\Delta^2/q^2$ and then putting $\Delta^2 =0$, i.e., replacing $d \rightarrow d^{PT}$ (for details see Section V and derivations therein). Now we want to replace the variable $z$
by the variable $z'=\Delta^2_t(d)/q^2 =c(d)\Delta^2 /q^2 = c(d) z$ and after the differentiation again to make the above-mentioned replacement. Then the expansion (10.11) will look like as

\begin{equation}
D_{\mu\nu}(q) =  { \Delta^2_t(d) \over (q^2)^2} i T_{\mu\nu}(q) \sum_{k = 0}^{\infty} \Bigl( {\Delta^2_t(d) \over q^2} \Bigr)^k \tilde{\Phi}_k(0),
\quad q^2 \rightarrow 0,
\end{equation}
because the additional factor $c^{-k}(d^{PT})$ will cancel the factor $c^k(d^{PT})$ in the initial coefficient functions (5.4), and that is why we replaced
$\Phi_k(0) \rightarrow \tilde{\Phi}_k(0)$. The renormalized gluon propagator (10.12) now becomes

\begin{equation}
\bar D_{\mu\nu}(q) = X'(\epsilon) \times { \bar \Delta^2 \over (q^2)^2} i T_{\mu\nu}(q) \sum_{k = 0}^{\infty} [X'(\epsilon)]^k
\Bigl( {\bar \Delta^2 \over q^2} \Bigr)^k \tilde{\Phi}_k(0), \quad \epsilon \rightarrow 0^+,
\end{equation}
due to the relation (10.2). From the expansion (10.5) to its leading order in the $\epsilon \rightarrow 0^+$ limit, one finally arrives at

\begin{equation}
\bar D_{\mu\nu}(q) \sim \epsilon \times A' \tilde{\Phi}_0(0) { \bar \Delta^2 \over (q^2)^2} i T_{\mu\nu}(q)
+ \bar O_{\mu\nu}(q; \epsilon^2), \quad \epsilon \rightarrow 0^+.
\end{equation}

It is instructive to get such kind of the expansion from the general expression (9.8). As in the case with the expansion (10.14) first necessary to go
to the $q^2 \rightarrow 0$. Then  to leading order in the $\epsilon \rightarrow 0^+$ limit, one finally obtains

\begin{equation}
D^R_{\mu\nu}(q) \sim \epsilon \times A Z_s { \Delta^2 \over (q^2)^2} i T_{\mu\nu}(q)
+  O^R_{\mu\nu}(q; \epsilon^2), \quad \epsilon \rightarrow 0^+,
\end{equation}
where $Z_s$ is the Picard constant (number), which appears due to the replacement $L(q^2) \rightarrow Z_s$ near small $q^2$ because of the Picard theorem,
see eq.~(5.22). The formal coincidence will be achieved by the finite re-scaling as follows: $A' \tilde{\Phi}_0(0) = A Z_s \bar{Z}^{-1}$ and $\bar \Delta^2 = \bar Z \Delta^2$. Let us remind that the combination $ A Z_s \Delta^2$, which appears in the expansion (10.15), has been denoted as
$\Delta^2_R = \Delta^2 Z_s A$ in the previous section.

The expansion (10.14) has been obtained from the expansion (10.13) by the suppression
of the higher order terms in the $\epsilon \rightarrow 0$ limit. At the same time, the expansion (10.15) has been obtained from eq.~(9.8) by the direct
$q^2 \rightarrow 0$ limit. However, the final expansions effectively coincide with each other. This means that  all the severe IR singularities apart from the simplest one $(q^2)^{-2}$ will be suppressed in both limits, in complete agreement with the Picard effect.
It requests the suppression of all the severe IR singularities apart from the simplest one in the $q^2 \rightarrow 0$ limit,
while our formalism developed here shows the same effect in the  $\epsilon \rightarrow 0^+$ limit. But the order of the limits to go is important. First the $q^2 \rightarrow 0$ limit in order to exactly and uniquely separate singular and regular parts, and only after the final $\epsilon \rightarrow 0^+$ limit, are to be taken, and not vice versa. As underlined above, such kind of the separation in the gluon multi-loop skeleton diagrams is to be carefully done as well.
The IR regularized solution (9.8) depends
on the mass gap, but remains massless, so the coincidence (up to numbers $A$ and $A'$) of the full gluon propagator (9.6) and the tadpole term INP MP IR renormalization (10.5) constants, respectively, is not surprising. The same effect occurs in the NP massive solution~\cite{35} when the mass gap goes to zero, and thus the massless solution is recovered. Then the gluon wave function renormalization constant coincides with mass gap renormalization one.
There is no doubt that effectively the IR renormalization properties of the full
gluon propagator and the tadpole term lead to the same results in the $\epsilon \rightarrow 0^+$ final limit. The explicit calculation of the tadpole term
(2.4) and its IR renormalization within our solution, described in Sections IX and X, is left for the readers (for this particular calculation is preferable to present eq.(10.2) as follows: $\tilde \Delta^2  = \tilde X(\epsilon) \Delta^2_t(d)$, and now $\tilde X(\epsilon) = \tilde{A} \epsilon + O(\epsilon^2), \ \epsilon \rightarrow 0^+$).

Concluding, let us note that within the mass gap approach to QCD and its phenomenological applications (see Appendixes B and C) always much more convenient to use the formalism based on the initial factorization (4.7). It has been developed in Section IX, where the mass gap
is the IR finite from the beginning, i.e., one can ignore the IR properties of the mass gap. This means that all the dependence on $\epsilon$
is attributed to the full gluon propagator. In many other applications
(see, for example~\cite{8} and references therein) one needs to consider the composition $(DD_0^{-1})$, which in the $q^2 \rightarrow 0$ limit does not contain
the severe IR singularities at all, but the dependence on the mass gap may remain. So that the IR properties neither the gluon propagator nor the mass gap
play any role, and they can be considered as IR finite, indeed. However, the formalism developed in this section is important from the theoretical
point of view. It explicitly  underlines the effective coincidence of the IR renormalization properties of the tadpole term with that of the gluon propagator itself when $\epsilon \rightarrow 0^+$.
It makes it possible to show the self-consistency of the Picard effect in both limits, first in the $q^2 \rightarrow 0$ limit
and than in the $\epsilon \rightarrow 0^+$ final one, etc.

\section{Discussion}
%\label{sec:sum}

One of the important conceptual problems in theoretical physics is the origin of a mass~\cite{58}, and hence the existence of the mass gap itself~\cite{9}. Our findings provide new insights into its dynamical generation at the fundamental quark-gluon level.
We have explicitly shown that the initial exact SU(3) color gauge symmetry of the QCD Lagrangian is not a symmetry of its ground state (vacuum) from the dynamical point of view, without mentioning its well-known topological complexities at the classical level~\cite{59} (instantons, monopoles, etc.). Quite possible that just due to our claim that the symmetries of the QCD Lagrangian and its ground state do not coincide, QCD is a self-consistent quantum field gauge theory. It needs no extra degrees of freedom in order to dynamically generate a mass.

Within our approach the gauge symmetry of the ground state has not been broken down by hand. By the design it is there from the very beginning.
It has been broken by the presence of the tadpole term in the dynamical and gauge structures of the full gluon SD equation, which describes the propagation of gluons in the QCD ground state. The tadpole term, having the dimensions of mass squared, is generated by the self-interaction of massless gluon modes, but
the point-like quartic gluon vertex is only involved at the skeleton loop level, see Fig. 1 and discussions in~\cite{8,9,60}.
However, the exact gauge symmetry requires its omission along with other mass scale parameters, having the dimensions of mass squared. So it plays no role in the preservation of this symmetry in the QCD vacuum.
In other words, the tadpole term is explicitly present in the QCD vacuum, but its true role has been hidden by the exact gauge symmetry. The derived splintering expressions (3.37) makes it possible to reveal the tadpole term, which renormalized version we call the mass gap, as the dynamical source of this symmetry breakdown in the vacuum of QCD. In this way we extend the concept of the mass gap to be also accounted for the QCD ground state.
Symbolically, it is possible to say that the mass gap has been 'closed' in the room and the splintering is a key to open it, and thus to disclose the
real role of the mass gap in the true dynamical and gauge structures of the QCD ground state. So that the splintering procedure permits to explicitly show up the mass gap in the full gluon SD equation and in the full gluon propagator as well, while respecting the corresponding ST identities for it and its free counterpart.

All this allows to formulate novel NP approach to QCD and its ground state in Section IV. We call it as the mass gap approach.
The general structure of the full gluon propagator, expressed in terms of the regularized quantities, has been derived and shown in eq.~(4.5).
Its longitudinal component is exactly defined as a function of $q^2$ and $\Delta^2_t(D)$, shown in eq.~(4.6). Its transverse component is
defined up to the invariant function $\Pi(q^2; D)$, which is regular at zero and may have the PT logarithmic divergences only: otherwise remaining arbitrary.
The full gluon invariant function (4.2) or, equivalently, (5.2) is nothing else but the transcendental equation.
If one formally ignores the tadpole contribution $\Delta^2_t(D)$, then we are left with the
PT massless full gluon propagator (3.21), and ignoring further the contribution from its invariant function, one arrives at the free
massless gluon propagator (3.3). The renormalization of the PT full gluon propagator is well-known procedure, see for example~\cite{3,4,5}, and it is not
our problem here, as emphasized above. Evidently, the PT massless gluon solution is due to the exact gauge symmetry preservation in the QCD ground state
(see the first subsection in Section III).

In close connection with the mass gap generation and gauge symmetry breakdown occurs the problem of the formulation of the corresponding renormalization theory. The tadpole term being the dynamical source of the gauge symmetry breakdown can not be renormalized within the PT method, since it is quadratically divergent constant, but regularized one. Due to the complicate NL transcendental character of the gluon invariant function (4.2), the renormalization theory, going beyond the PT technics, depends on its concrete solutions. It may have the different solutions.
Keeping the tadpole contribution 'alive', one comes to the two different types of the general solutions: the INP singular full gluon propagator (9.1)
derived and renormalized here in Sections V-X. The NP massive but regular full gluon propagator derived and renormalized in~\cite{35}.
They are resulting of the gauge symmetry violation in the QCD ground state. For the first solution the corresponding INP MP IR renormalization
program has been formulated in order to make the theory free from all the types of the quadratic UV divergences and severe IR singularities at the final stage.
It is based on the joint use of the theory of functions of complex variable, the theory of distributions (generalized functions) and DRM, correctly implemented into this theory. For the second solution the corresponding NP MP renormalization program has been performed which made the theory finite up to the PT renormalizable logarithmic divergences only.
Despite the gauge symmetries of the QCD Lagrangian and its ground state are not the same,
nevertheless, the renormalizability of the QCD/YM theory is not affected, i.e., it remains renormalizable within the mass gap approach to the QCD.
May be it sounds as a paradox, but it seems to us that the INP renormalizazion program described here is not so cumbersome as the PT renormalization program itself. Apparently, it is due to a firm mathematical background behind the former one.

The important issue of the renormalization beyond the PT method within the mass gap approach to QCD needs a few clarifying remarks in addition.
For the renormalization of the PT QCD it is essential to remove from the theory the quadratically divergent scale parameters
since it fails to renormalize them. The exact gauge symmetry makes it possible to achieve this goal. For the NP approach the presence
of the quadratically divergent scale parameters (which violet the exact gauge symmetry) is not a problem, as it has been pointed out just above.
Briefly speaking, the mass scale parameters
(the quadratically divergent from above or severely singular from below) being an essential problems for the PT QCD, are not the problems
for our general approach, i.e., being unrenormalizable in the PT QCD such quantities can be renormalized in the NP/INP QCD.
In other words, in the PT QCD the exact gauge symmetry can not be broken in its ground state. In the NP QCD and in its INP phase
the exact gauge symmetry has to be broken, otherwise impossible to explain color confinement and the scale violation in the AF regime, as it has been done
in the present work. There is no doubt that the dynamical and gauge structures of the PT QCD and NP/INP QCD ground states are substantially different.
Let us emphasize that when we are speaking (above and below) that the QCD Lagrangian's gauge symmetry is broken down or violated or not
the same, etc. as in its ground state,
we mean that the terms (having the dimensions of mass squared) which cannot be present in the Lagrangian may still present in the ground state.
Moreover, they can survive the corresponding renormalization program if it goes beyond the PT technics, i.e., such kind of programs (depending on the above-mentioned solutions) are to be essentially NP. Precisely in this sense the renormalizability of QCD/YM as a theory of strong interactions
is to be understood.

In order to complete these general remarks on the NP/INP renormalization, it is necessary to emphasize the important role of the ST identity
for the gauge particles in the renormalization program beyond the PT methods. As underlined above, the ST identities are consequences of the exact
gauge invariance/symmetry in the PT sense. However, this symmetry is broken in the QCD/YM ground state, and it is accompanied by the gauge-changing phenomenon.
Precisely to it the ST identity's role in the NP renormalizability of the theory
remains crucial within the mass gap approach to QCD. The INP MP IR renormalization of the gauge particle ST identity requires that the
IR renormalization constants for the gluon wave function (or, equivalently, for the full gluon propagator in this case) and the tadpole term are effectively
the same. Otherwise, the general IR renormalizability of the theory would be destroyed. Also, it provides an exact constraint on the solutions to QCD within the mass gap, keeping it 'alive' in its ground state. Our main conclusion is that the role of the ST identity remains important whether the exact gauge symmetry is preserved or not.
In the first case it ensures the renormalizabilty of the PT, while in the second case it ensures the renormalizability beyond the PT.
In the IR renormalized singular solution the gauge-fixing term (which becomes the function of $q^2$) vanishes at
large distances ($q^2 \rightarrow 0$). But, nevertheless, the IR renormalization of the ST identity, which determines this term, remains essential, as pointed out above (see Sections IX and X).

The last conceptual problem solved within the mass gap approach is that the transverse gluons ('dressed' or free) can not appear as physical states
at large distances by any means. The confining full gluon propagator or, equivalently, the INP full gluon propagator dominates over all
the other possible solutions for full gluon propagator at $q^2 \rightarrow 0$, and in this limit it becomes transverse. This dominance
is a gauge-invariant result in the $\epsilon \rightarrow 0^+$ limit as well, and automatically provides confinement of color gluons.
By product to the solution of color gluons confinement
problem, we conclude that within the mass gap approach to QCD, it effectively becomes the abelian-type theory in the form briefly described
at the end of Section IX. This will be clearly seen by going beyond the YM sector.

Let us remind that the AF phenomenon consists of the two conditions. First necessary -- the scale violation, and second sufficient-- the positive sign
of the color charge interactions (for the purely YM theory it is always positive, while for the QCD it remains positive for the number of
different quark flavors less than 16). The AF has been discovered long ago~\cite{61,62}, but the scale violation remained mysterious since there.
As we already know, the PT is not able to generate a mass scale parameter by itself.
The renormalization group equations formalism assumes its existence~\cite{2,3,4,5,39} but where it comes from? remained an open question up to present days.
The Picard effect within the mass gap approach to QCD  proves its existence, and thus naturally explains this long-standing problem, namely how
the finite mass scale parameter appears under the PT logarithm. In fact, $\Lambda^2_{YM}$ is nothing else but $\Delta^2_{PT}$, defining in the relation
(5.23) and surviving the PT regime. It comes from the deep IR region and being thus of the INP dynamical origin.

In the presence of the mass gap the coupling constant $g^2$ plays no any role. This is also evidence of the 'dimensional transmutation',
$g^2 \rightarrow \Delta^2$~\cite{2,63,64}, which occurs whenever a massless theory acquires mass dynamically. It is a general feature of spontaneously symmetry breaking in field theories. Therefore, we distinguish between the PT and NP QCD by the explicit presence of the mass scale parameter -- the mass gap -- in the latter one, and not by the magnitude of the coupling constant. The INP QCD additionally distinguishes from the NP QCD by the explicit presence of severe IR singularities in the former one. This makes it possible to calculate such vacuum characteristics as the bag constant, the gluon condensate, the topological
susceptibility, etc.~\cite{8,45,46,47,48,49,50,51} (and references therein as well as see discussion in the Appendix C) in a self-consistent way. The problem is that all such quantities should be free from all the types of the PT contributions ("contaminations"). In our case of the INP singular solution (9.1)-(9.2) for the full gluon propagator the gluon remains massless, but, nevertheless, the propagator itself depends on the mass gap. At the same time, its PT part (which includes its longitudinal component) vanishes at large distances ($q^2 \rightarrow 0$) at any gauge. Thus the surviving part (which is nothing but the INP term in the full gluon propagator) is not "contaminated" by the PT contributions, indeed. Moreover, it becomes automatically transverse as well. It seems that we need no ghosts in this case. However, they will come into the play again in the PT $q^2 \rightarrow \infty$ limit. The dependence on the mass gap is suppressed in the longitudinal
component of the full gluon propagator in this regime, and thus its gauge-fixing parameter becomes $\xi_0=\lambda^{-1}$.

Though the PT massless full gluon propagator (3.21) has been previously called as a hypothetical one, nevertheless, it should be equally taken into account
along with its free massless counterpart. So that within the mass gap approach, one can formally register the four different general gluon 'flavors',
since no any appoximations/truncations/assumptions have been made in the derivation of these gluon 'species'.
At the same time, any correct concrete solutions for the PT full gluon propagator (3.21) are subject to the different appoximations/truncations/assumptions imposed on its invariant function.
All the three non-trivial solutions for the full gluon propagator and its free massless counterpart are valid in the whole gluon momentum range from zero to infinity, i.e, at all distances. However, the INP singular one dominates over all the other ones at large distances. Therefore just its renormalized version (9.8) correctly describes the transition to the confining phase in the YM theory. Moreover, the renormalization properties of the mass gap as a function of the dimensional IR regularization parameter $\epsilon$ also confirms its role in this transition in the region of small $q^2$ (see Section X).
No mass gap -- neither confining phase nor AF phenomenon in the YM theory. Quark/gluon confinement and AF, being experimental facts, have to be considered
as the 'boundary conditions' at $q^2 \rightarrow 0$ and $q^2 \rightarrow \infty$, respectively,
for any solution to QCD/YM at the fundamental quark-gluon level. Only the INP singular solution found here satisfies to gluon confinement as well as AF, while leading to the finite result (9.7) when the gluon momentum is an independent loop variable. In this connection,
it is worth underlining that the system of eqs.~(3.18)-(3.21), which preserves the exact gauge symmetry of the QCD Lagrangian, can explain neither
color confinement nor AF, since it does not possess the mass gap. Therefore, its solution does not satisfy to the above-mentioned 'boundary conditions'.
In other words, both effects/phenomena (color confinement and AF) are bright evidences of the QCD Lagrangian's gauge symmetry
breakdown in its ground state by the mass gap. Just this is reflected by the derived system of eqs.~(3.37)-(3.39).
We hope that we have conveyed this message to the readers as much clearly as possible.

Our mass gap coincides with the JW's one~\cite{9} by properties, but not by definition. Their definition requires that the Hamiltonian of the theory
has no spectrum in the interval $(0, \Delta)$, i.e., every excitation of the vacuum has energy at least $\Delta$. The supremum of
such $\Delta$ has been called the mass gap. In our case every excitation of the vacuum has energy which is to be
necessarily expressed in terms of the mass gap $\Delta$, but not only being at least $\Delta$. The main structural dynamical element of the full gluon propagator within the mass gap approach is the ratio $\Delta^2 / q^2$. The massive excitations investigated in~\cite{35}
has energy at least $\Delta$, indeed, where  $\Delta^2$ is exactly defined gluon pole mass squared $M_g^2$. So that the above-mentioned ratio is
$q^2 = M^2_g$, since the full massive gluon propagator behaves as $D_{\mu\nu}(q) \sim 1 / (q^2 - M^2_g)$ near the pole position.
At the same time, the vacuum excitations, investigated in the present work, being massless, but depending on the mass gap, may have energy of any magnitude. At short distances $q^2 \rightarrow \infty$ the ratio becomes $(\Delta^2 / q^2) < 1$. Then the energy of the vacuum excitations in the AF regime is to be expressed in terms of $\Delta_{PT} = \Lambda_{YM}$, and thus the ratio becomes $q^2 >  \Lambda^2_{YM}$. The value $\Lambda^2_{YM}=  0.09 \ GeV^2$~\cite{40} can
be treated as the scale of the non-trivial PT dynamics in the YM vacuum.
At large distances $q^2 \rightarrow 0$ the ratio becomes $(\Delta^2 / q^2) > 1$. Then the energy of the vacuum excitations in this low-energy limit  is to be expressed in terms of $\Delta_{JW}$, and thus the ratio becomes $\Delta^2_{JW} > q^2$ (let us remind that at the level of a single full gluon propagator
$\Delta^2_{JW}$ has been denoted as $\Delta^2_R = \Delta^2 Z_s A$).
The scales of the INP and NP dynamics are not identical, but can be of the same order of magnitude. As it follows from our estimates, comparable with lattice results (see Appendix B) and from~\cite{35}, one gets $\Delta^2_{JW} \sim M_g^2 \sim (0.2 - 0.3) \ GeV^2$.
Let us underline that the energy of the gluonic excitations of the YM vacuum, such as mentioned above the bag constant, gluon condensate, topological
susceptibility, etc.~\cite{8, 45,46,47,48,49,50,51} is to be measured in terms of the JW mass gap $\Delta_{JW}$ in the corresponding powers
(see also Appendixes B and C).

The general expression (4.5) obtained for the full gluon propagator within the mass gap approach to QCD is the NL transcendental equation.
We have shown that keeping the mass gap 'alive' it has the two general independent solutions: the NP massive~\cite{35}, which is regular in the
the $q^2 \rightarrow 0$ limit, and the INP massless, investigated here, which is singular in this limit. Putting formally the mass gap zero
in eq.~(4.5), it has the two massless solutions, the PT gluon propagator and free gluon one. Let us now perform some rather simple exercise, namely
by going to the $q^2 \rightarrow 0$ limit directly in eq.~(4.5). Remembering that its invariant function is regular at zero, one finally obtains

\begin{equation}
D_{\mu\nu}(q) =   i \delta_{\mu\nu}{ 1 \over \Delta^2_t(D)}, \quad q^2 \rightarrow 0.
\end{equation}
This asymptotic coincides with the asymptotic of the regular massive solution, but completely contradicts to the IR asymptotic of the singular solution (5.14). In other words, the INP massless solution will be missing in this procedure. As pointed out above, going to the $\Delta^2=0$ limit, both solutions, depending on the mass gap, will be missing as well. In the PT $q^2 \rightarrow \infty$ limit the explicit dependence on the mass gap vanishes
in eq.~(4.5), and thus it solution will be reduced to the PT massless one. However, the dependence on the mass gap will appear under the PT logarithms
within the singular solution in this limit, as explained in Section 7. All these inconsistences
happened because in the derivation of the expression (11.1), we have completely ignored the NL transcendental character of the gluon invariant function,
associated with the transverse component in the general expression (4.5). We treated it in the same way as the standard function of $q^2$, associated with its longitudinal component. These simple derivation and associated with it arguments, reflect the general situation with the NL transcendental equations. Namely,
it is always dangerous to go directly to some limits with respect to the variables and parameters in them. This is also true if one introduces into the NL transcendental expressions any kind of the approximations/truncations/assumptions. The introduction of such kind of the simplifications (including the linearisation) may violate the NL transcendental character of the initial expressions. As a result, one may loss a lot of useful information about the true
dynamical and gauge structures of the QCD ground state, reflected by its SD system of equations of motion. The inial gluon SD eq.~(2.1) is a such NL
transcendental one from the very beginning. Our gluon SD eq.~(4.5) does not destroy the cluster~\cite{9} (i.e., its NL and transcendentality)
character of its skeleton loop integrals (shown in Fig. 1), since we do not introduce in to them any kind of the simplifications, indeed.
In eq.~(4.5) these skeleton loop integrals are save and present simply in other form. We consider this as a big advantage of the mass gap approach to QCD.
That was the main reason why we were able to find the two independent general solutions, depending on the mass gap in the two different ways.

The purely YM theory with the preservation of the exact gauge symmetry/invariance looks like the conformal field theory (CFT)~\cite{65}, i.e., it can be
considered in some sense as the conformal gauge theory.  The former one forbids the existence of the finite mass squared parameters, while
the latter one has no finite specific length scale, which has the dimension of inverse mass in natural units.
In other words, both theories have no any finite parameters which dimensions can be expressed in terms of a non-zero finite mass.
From this point of view, apparently, one can speak in terms of the scale/conformal invariance breakdown in the YM theory within our approach, indeed.
However, its connection (if any) to the boundary CFT (see recent work~\cite{66} and references therein), in which the boundary might be treated as the existence of a finite characteristic length scale, is not clear for us. How such a scale is to be dynamically generated in the boundary CFT?,
or is it already there due to its initial lattice regularization~\cite{66}?
Another possible way of the interpretation of our results is to consider YM Lagrangian, forbidding the presence of the parameters, having the
dimensions of mass squared, as the Lagrangian providing a 'chiral' symmetry of gluon interactions. This symmetry is broken down in its
ground sate, where the two different gluon 'flavors' appear, depending on the mass gap and mentioned above.  Of course, such 'chiral' gluon
symmetry and its break down is different from the chiral quarks flavor symmetry breakdown in the QCD Lagrangian.

\hspace{2mm}

Our the most general conclusion is: the PT is not applicable to QCD at all. In this paper we have shown its failure in the low-energies region
as well as even at high energies, because the PT can not explain the scale violation in the AF regime. At finite energies its failure
has been shown in detail in our work~\cite{35}. It is time already to forget about the PT way of thinking in dealing with QCD, only the NP methods are to be used for this purpose.

\hspace{2mm}

\section{Summary}

Let us now summarize our results in detail. These are independent, of course, from the comparison with any other possible
interpretations, for example briefly discussed just above in the last absatz of the previous section.

\begin{itemize}
\item[(i).] We have investigated the transversity of the full gluon self-energy in the way not using the PT, but using only the corresponding ST identities
for the full and free gluon propagators.
\item[(ii).] We extend the concept of the mass gap to be accounted for the QCD ground state (vacuum) as well.
\item[(iii).] The dynamical source of the mass gap has been identified with the tadpole/seagull term, which is explicitly present in the full gluon self-energy. It is generated by the self-interaction of massless gluon modes.
\item[(iv).] We have demonstrated that the exact gauge symmetry of the QCD Lagrangian is dynamically broken down by the explicit presence of the mass gap
in its ground state. This violation is to be accompanied by the gauge-changing phenomenon $\xi \neq \xi_0$.
We have also explain how the exact gauge symmetry of the QCD Lagrangian might be preserved in its vacuum, which leads to the relation $\xi = \xi_0$.
\item[(v).] The splintering expressions (3.37) for the corresponding transverse conditions has been derived. It made it possible to show up the mass gap
in the gluon SD equation, and thus in the full gluon propagator.
\item[(vi).] On this basis the mass gap approach to QCD has been formulated. It shows the general structure of the full gluon propagator (4.5).
\item[(vii).] It uniquely implies the generalized gauge (4.6), derived in new manner as a function of the mass gap and the gluon momentum $q^2$.
\item[(viii).] We have derived the general INP singular solution (9.1) for the full gluon propagator. It summarises all the severe (i.e., INP) IR
singularities when the gluon momentum goes to zero, due to the non-abelian character of QCD. Then all the QCD/YM vertexes can be effectively
considered as regular functions of their gluon momenta involved.
\item[(ix).] All the INP IR singularities are associated with the transverse component of the full gluon propagator.
They cannot be summed up to a some known function. Its longitudinal component becomes known regular function of the gluon momentum,
see eqs.~(4.6) and (5.16).
\item[(x).] No any appoximations/truncations/assumptions have been made in the derivation of our solution. The dependence on all the
un-physical parameters disappears after the corresponding renormalization program is performed.
\item[(xi).] The INP part of the singular solution (5.14) has been found in closed form (5.11)-(5.13). The corresponding Laurent series
is the cluster expansion~\cite{9}, where the dependence on the coupling constant $g^2$ is not simple. Moreover, due to the Picard theorem
the convergence of our solution to the desired finite limits has been established.
\item[(xii).] We have demonstrated that QCD as a quantum field gauge theory with the explicit presence of the mass gap in its ground state
possess essential singularities in the limit of very small and very large
gluon momenta. It has been done in the framework of the INP singular solution for the full gluon propagator and its asymptotics.
\item[(xiii).] We have proposed the formalism how to correctly treat the severe IR singularities. This has been achieved by
combining the theory of distributions (generalized functions), the theory of functions of complex variable and the DRM in a self-consistent way.
For the first time, it has been formulated in some details in~\cite{8}.
\item[(xiv).] We have discovered that the composition $\Delta^2/(q^2)^2$ plays the important role in the dynamical structures of the QCD
ground state. According to the Picard theorem only the simplest severe IR singularity $\sim (q^2)^{-2}$ survives in a single full gluon propagator
in the $q^2 \rightarrow 0$ limit, indeed.
\item[(xv).] We have explicitly shown the existence of the two different finite mass scale parameters on the basis of the two different
renormalization programs of the same mass gap have been performed. They are closely related to the above-mentioned QCD essential singularities.
\item[(xvi).] The first one possesses
the physical meaning of the scale responsible for the INP confining dynamics in the QCD ground state at large distances.
We call it as the JW mass gap $\Delta^2_{JW}$. The second one possess the physical meaning of the scale responsible
for its non-trivial PT AF dynamics at short distances, $\Delta^2_{PT} = \Lambda^2_{YM}$.
\item[(xvii).] This made it possible to explain the linear rising potential between heavy quarks as well as the scale violation in the AF regime.
For this the running effective charge in the corresponding asymptotics is needed, depending on the mass gap.
\item[(xviii).] We establish that the QCD/YM is a spontaneously/dynamically broken gauge theory in its ground state, but remains renormalizable
within our approach, while respecting the corresponding ST identities. It has a correct PT $q^2 \rightarrow \infty$ limit with
the scale violation in the AF regime.
\item[(xix).] The INP MP IR renormalization program has been developed within the mass gap approach.
It is based on the newly-derived DRE (8.10). It shows that all the severe IR singularities scale as $1/ \epsilon$ only at $\epsilon \rightarrow 0^+$.
\item[(xx).] We have found that the IR renormalization of the ST identity is important, whether the exact gauge symmetry is preserved or not.
It requires that the IR renormalization constants of the gluon wave function and the tadpole term have to be effectively the same: otherwise the general
IR renormalizability of the theory would be ruined.
\item[(xxi).] When the gluon momentum is an independent loop variable the solution for the renormalized full
gluon propagator (9.7) is transverse and uniquely fixed as a function of the gluon momentum.
\item[(xxii).] The general expression for the INP renormalized full gluon propagator (9.8) is fixed. It is valid in the whole gluon momentum range.
\item[(xxiii).] The massless gluons (but depending on the mass gap) can not appear at large distances as physical states, see eq.~(9.9).
All other types of gluons are suppressed at this distances (confinement of color gluons).
\end{itemize}

\hspace{2mm}

In short, the resume of our main results is as follows: the discovery of the true dynamical and gauge structures of the QCD vacuum
makes it possible to explain AF, color gluons confinement, linear rising potential, etc. The crucial role in this discovery
has been played by the extension of the JW mass gap concept~\cite{9}, first introduced within the Hamiltonian formalism of QCD, to be accounted for its ground state as well. As explained above, our mass gap coincides with the JW's one by properties, but not by definition.
Within our approach the mass gap has a multiple 'faces'/interpretations/meanings. At the regularized level, linearly contributing to the full gluon self-energy as the tadpole term, it possesses the meaning of the scale, which violates the exact gauge symmetry of the QCD Lagrangian in its ground state.
NL appearing in the full gluon propagator (4.5), it possesses the meaning of the scale, which separates its solutions, depending on the tadpole term from
those not depending on it at all. In the massive solution~\cite{35} its renormalized version becomes the exactly defined gluon pole mass, and thus
determines the scale of the NP dynamics in the QCD vacuum at finite distances. In the singular solution its renormalized version can be interpreted even in the two different ways: it becomes the JW scale, which determines the confining dynamics in the QCD vacuum at large distances. At short distance it becomes the scale, which determines the scale of its non-trivial AF dynamics (see Section VII).

The important challenges, formulated in Section I, have been addressed and solved by the mass gap approach to QCD ground state.
By solutions, we mean first of all the formulation of the corresponding NP and INP renormalization programs maintaining the renormalazibility of QCD
in the explicit presence of the quadratically divergent constants and severely singular quantities, and thus remaining unrenormalizable by the PT technics.
Also, it allows to determine the asymptotical and analytical properties of the gauge particles (gluons) Minkowskian and Euclidean full
Green's functions. The former ones describe the propagation of the massive gluons investigated in~\cite{35}. The latter ones
describe the propagation of the severely singular massless gluons investigated in the present work.

As a subject for further perspective work is the quark confinement phenomenon. The confining full gluon propagator derived here can be a key
to the solution of this important conceptual problem. The very preliminary results are impressive, but, unfortunately, a long-term work is awaiting for.

Finally, we note that (unlike the ghost and tadpole terms) the gauge-fixing term in the QCD Lagrangian does not explicitly contribute to the full gluon self-energy and thus to the full gluon SD equation and its solutions, obtained within the mass gap approach to the QCD ground state.
So that we think that its formulation does not depend on this term. That is why it has been left out of our consideration.

\begin{acknowledgments}

The authors are grateful to P. Forg\'{a}cs, J. Nyiri, T.S. Bir\'{o}, M. Vas\'{u}th, Gy. Kluge
for useful suggestions, remarks, discussions and help.
The work was supported by the Hungarian National Research, Development and Innovation Office (NKFIH) under the contract numbers OTKA K135515, K123815 and NKFIH 2019-2.1.11-T\'ET-2019-00078, 2019-2.1.11-T\'ET-2019-00050, the Wigner GPU Laboratory and THOR Cost Action CA15213.

\end{acknowledgments}

\appendix

\section{ The non-splintering procedure }

The splintering procedure in respect with how to explicitly show that the exact gauge symmetry/invariance
in the QCD ground state is already broken from the very beginning has been described in Section III. The only other possible way
to do this is presented in this Appendix. As we already know from the relations (3.5) and (3.8) in the general case
\begin{equation}%\label{eq:a1}
q_{\rho} q_{\sigma} \Pi_{\rho\sigma} (q; D) \neq 0,  \quad \quad      q_{\rho} q_{\sigma} \Pi^{(s)}_{\rho\sigma} (q; D) \neq 0,
\end{equation}
i.e., both transverse conditions are not satisfied, which means there is no the splintering (3.37) at all. However, in the similar way exploited in Section III, let us begin with the corresponding subtraction
\begin{equation}
\Pi^{(s)}_{\rho\sigma}(q; D)= \Pi_{\rho\sigma}(q; D) - \Pi_{\rho\sigma}(0; D)= \Pi_{\rho\sigma}(q; D) - \delta_{\rho\sigma} \Delta^2(D),
\end{equation}
with $\Pi^{(s)}_{\rho\sigma}(0; D) =0$,
and where by $\Delta^2(D)$ we denote the sum which appears in eq.~(3.7), namely $\Delta^2(D) = \Delta^2_q + \Delta^2_g(D) + \Delta^2_t(D)$.
The general decompositions of the full gluon self-energy and its subtracted counterpart into the independent tensor structures are given in
the relations (3.9), namely
\begin{eqnarray}
\Pi_{\rho\sigma}(q; D) &=&  T_{\rho\sigma}(q) q^2 \Pi_t(q^2; D) - q_{\rho} q_{\sigma} \Pi_l(q^2; D), \nonumber\\
\Pi^{(s)}_{\rho\sigma}(q; D) &=&  T_{\rho\sigma}(q) q^2 \Pi^{(s)}_t(q^2; D) - q_{\rho} q_{\sigma} \Pi^{(s)}_l(q^2; D).
\end{eqnarray}
The both invariant functions $\Pi^{(s)}_t(q^2; D)$ and $\Pi^{(s)}_l(q^2; D)$ cannot have power-type singularities (or, equivalently, pole-type ones) at small $q^2$, since $\Pi^{(s)}_{\rho\sigma}(0; D) =0$ by definition in eq.~(A2): otherwise they remain arbitrary. On account of the subtraction (A2) and the relations (A3), one obtains
\begin{eqnarray}
\Pi_t (q^2; D) &=&  \Pi^{(s)}_t(q^2; D) + {\Delta^2(D) \over q^2}, \nonumber\\
\Pi_l(q^2; D) &=&   \Pi^{(s)}_l(q^2; D) - {\Delta^2(D) \over q^2},
\end{eqnarray}
then the full gluon self-energy becomes
\begin{equation}
\Pi_{\rho\sigma}(q; D) = T_{\rho\sigma}(q) \left[ q^2 \Pi^{(s)}_t(q^2; D) + \Delta^2(D) \right]
- L_{\rho\sigma} \left[ q^2 \Pi^{(s)}_l(q^2; D) - \Delta^2(D) \right].
\end{equation}
Substituting eq.~(A5) into the initial gluon SD eq.~(2.1), one arrives at

\begin{eqnarray}
D_{\mu\nu}(q) &=& D^0_{\mu\nu}(q) + D^0_{\mu\rho}(q)i T_{\rho\sigma}(q) \left[ q^2 \Pi^{s}_t(q^2; D) +  \Delta^2(D) \right] D_{\sigma\nu}(q) \nonumber\\
&-& D^0_{\mu\rho}(q)i L_{\rho\sigma}(q) [ q^2 \Pi^{(s)}_l(q^2; D) - \Delta^2(D)] D_{\sigma\nu}(q).
\end{eqnarray}
Contracting both sides of this equation with $q_{\mu}$ and $q_{\nu}$ and doing some algebra using the decompositions (3.2)
and (3.3), one finally gets

\begin{equation}
\xi \equiv \xi(q^2) = { \xi_0 q^2 \over q^2 + \xi_0 [\Delta^2(D) - q^2 \Pi^{(s)}_l(q^2; D)] },
\end{equation}
instead of eq.~(4.4). Let us remind that in this expression $\Delta^2(D) = \Delta^2_q + \Delta^2_g(D) + \Delta^2_t(D)$, while in
the solution (4.4) it is the tadpole constant itself, i.e.,  $\Delta^2_t(D)$.
If both transverse conditions (A1) were satisfied, i.e., equal zero, then $\Delta^2_q = \Delta^2_g(D) = \Delta^2_t(D) =0$, and
we would have reproduced the correct limit of the exact gauge symmetry.

The last expression (A7) in the PT $q^2 \rightarrow \infty$ limit becomes

\begin{equation}
\xi \equiv \xi(q^2) = { \xi_0  \over 1 - \xi_0 \Pi^{(s)}_l(q^2; D) },
\end{equation}
which is equivalent to the formal $\Delta^2(D) = 0$ one. However, as we already know, the arbitrary invariant function $\Pi^{(s)}_l(q^2; D)$ may have logarithmic divergences
at $q^2 \rightarrow \infty$. So that $\xi(q^2) \rightarrow - 0$ in this limit, while in the generalized gauge (4.4) it goes to
$\xi(q^2) \rightarrow \xi_0$, which is only one correct. In order to get the correct limit from the general expression (A7) one needs to put
$\Pi^{(s)}_l(q^2; D) =0$ there. Then from the relations (A3) and (A4) one finally arrives at the splintering relations (3.37), indeed.
This is one more strong argument in favor of our formalism, developed in Section III, in order to prove that the gauge symmetries of the
QCD Lagrangian and its ground state are not the same. Let us remind that the second of the transverse relations (3.37) requires
$\Delta^2_q = \Delta^2_g(D) = 0$, while $\Delta^2(D) = \Delta^2_t(D)$ remains finite.

On the other hand, combining eq.~(A5) and the general transverse condition (3.5), one obtains

\begin{equation}
[\Delta^2(D) - q^2 \Pi^{(s)}_l(q^2; D)] = \left( {\xi_0-\xi \over \xi \xi_0} \right) q^2,
\end{equation}
and substituting it back to the previous eq.~(A7), one arrives at the identities $\xi= \xi$ as well as $\xi_0 = \xi_0$.
This is in complete agreement with the identities which come from the initial gluon SD eq.~(3.12) contracted with $q_{\mu}$ and $q_{\nu}$.
We consider these identities as a test on the correctness and uniqueness of the mass gap approach to QCD.
They confirm that the gauge choice by the relation (3.36) and the equivalent solution (4.4) have been justified.
Only the expression (4.4) has a correct PT $q^2 \rightarrow \infty$ limit, which is equivalent to the formal $\Delta^2_t(D) = 0$ one, as it was described
in Section IV. The unique way to correctly show up the mass scale parameter in the full gluon propagator
is the splintering expressions (3.37), explicitly based on the tadpole term as the mass gap.

\section{The linear rising potential}
%\label{euclid}
%\label{sec:6}

At large distances ($q^2 \rightarrow 0$) the full gluon propagator (5.14) is dominated by its INP part (6.1)-(6.2). It is nothing else but the
summation of all the possible INP IR singularities, which may appear in QCD ground state due to the self-interaction of  massless gluon modes.
However, because of the Picard effect, discussed
in detail in Sections V, VI and IX, survives only the simplest INP IR singularity, so that the full gluon propagator at large distances
is dominated by its expression (5.20), namely

\begin{equation}
D^{INP}_{\mu\nu}(q) = i T_{\mu\nu}(q) { \Delta_{JW}^2 \over (q^2)^2}, \quad q^2 \rightarrow 0,
\end{equation}
where instead of $\Delta^2_R$ (see Section X) we equivalently denote $\Delta^2 Z_s A = \Delta^2_{JW}$.
The full gluon propagator does not depend
on any un-physical parameters, i.e., it becomes renormalized. The corresponding running effective charge is

\begin{equation}
\alpha(q^2; \Delta_{JW}^2) = { \Delta_{JW}^2 \over q^2}.
\end{equation}

The potential between heavy quarks is defined as follows:

\begin{equation}
V(r) = - 4 \pi C_2(R) \int { d^3 q \over (2 \pi)^3} e^{i \bar{q} \bar{r}} { \alpha(q^2; \Delta^2_{JW}) \over q^2},
\end{equation}
where $C_2(R) = (N_c^2 -1)/ 2N_c = 4/3$ is the positive eigenvalue of the quadratic Casimir operators for the fundamental representation of $SU(3)$.

The general expression for the power-type effective charges is~\cite{42}

\begin{equation}
\int { d^n q \over (2 \pi)^n} e^{iqr} { 1 \over (q^2)^{\lambda}} = { \Gamma(n/2 - \lambda) \over 4^{\lambda} \pi^{n/2}
\Gamma(\lambda)} (r^2)^{\lambda - n/2}.
\end{equation}
Substituting further eq.~(B2) into eq.~(B1) and using this expression for $\lambda =2, \ n=3$ and $\Gamma(-1/2)= - 2 \sqrt{\pi}$,
one finally obtains

\begin{equation}
V^{INP}(r) = { 2 \over 3} \Delta^2_{JW} \ r = \sigma \ r, \quad r \rightarrow \infty,
\end{equation}
and $\sigma$ is the string tension. The relation between the scale of the INP QCD $\Delta^2_{JW}$ and $\sigma$ then becomes

\begin{equation}
\Delta^2_{JW} = { 3 \over 2} \sigma.
\end{equation}

From lattice simulations~\cite{40,43,44} it is well-known that

\begin{equation}
\sqrt{\sigma} = 0.420 \ GeV, \quad \sigma = 0.1764 \ GeV^2,
\end{equation}
so that for the scale of the INP dynamics in QCD, one obtains the value as follows:

\begin{equation}
\Delta_{JW} = 0.5144 \ GeV, \quad  \Delta^2_{JW} = 0.2646 \ GeV^2.
\end{equation}
Of course, this value should not be accepted literally, since it depends on the string tension's value, but the relation (B6) is exactly defined.

Is is instructive to calculate the potential for the PT gluon propagator (5.17), which dominates the full gluon propagator at very large gluon momentum
(short distance), but nevertheless formally can be considered in the whole gluon momentum range. At $q^2 \rightarrow 0$, i.e., at large distance, its effective charge becomes simply constant, namely $d^{PT}(0) = \alpha^{PT}(0)$, which for the free gluon effective charge is $d^0(0) = \alpha^0(0)=1$. Then from the expressions (B1) and (B3) for $\lambda =1, \ n=3$ and $\Gamma(1/2)= \sqrt{\pi}$, one finally obtains the Coulomb potential, as expected

\begin{equation}
V^{PT}(r) = - { 4 \over 3} { \alpha^{PT}(0) \over r}, \quad r \rightarrow \infty,
\end{equation}
and thus the full Cornell potential becomes

\begin{equation}
V(r) = V^{INP}(r) +V^{PT}(r) = { 2 \over 3} \Delta^2_{JW} \ r - { 4 \over 3} { \alpha^{PT}(0) \over r}.
\end{equation}
Let us point out that at finite gluon momenta, i.e., at small but finite distances, the potential is of Yukawa-type, namely
$V^{NP}(r) \sim (1/r) \exp(-M_gr)$, where $M_g$ is the exactly defined gluon pole mass~\cite{35}. So that at very large and very small distances
the Cornell potential (B10) is dominated by its INP and PT contributions, respectively, while at the finite distances its NP contribution has to be taken
into account as well.

Concluding, the Picard effect is a main "culprit" behind the phenomenon that the potential between heavy quarks is only linear confining one,
calculating by lattice QCD as well~\cite{43,44}, as pointed out above, i.e., there is no any power-type confining potential at large distances.
Due to this effect only simplest INP IR singularity $\sim 1/ q^4$ survives at large distances ($q^2 \rightarrow 0$), while all others are suppressed. This is the one of main phenomenological consequences which directly follows from the mass gap approach to the QCD ground state. If its gauge symmetry would have been the same as of the QCD Lagrangian, then there were no the linear rising potential at all, and only Coulomb-type one would have been existed.
In other words, the Picard effect, which shows up itself within the mass gap approach, explains why there is only linear rising potential between
heavy quarks in QCD.

And finally, to analyse the interaction between heavy quarks in terms of the confining linear rising potential is justified. In this
case the response of the vacuum (radiative corrections or, equivalently, vacuum fluctuations) can be neglected, so that the full quark-gluon vertex is well-approximated by its point-like counterpart. For light quarks this simplification does not work, since the response of the vacuum is strong, and the full quark-gluon vertices are to be taken into account. So the only way to understand confinement of light quarks is the investigation of analytical and asymptotical properties of the quarks Green's functions, i.e.,  their full propagators and the corresponding full vertices. They will be substantially
modified due to the strong vacuum response. It has been shown~\cite{67,68} that the ladder approximation (LA) is not self-consistent in QCD due to the strong interaction between color charges. It can be used only for heavy quarks, as explained above, or in the AF regime, when the coupling constant becomes weak.

\section{The renormalized running effective charge and the $\beta$-function}

%\vspace*{-1.0truecm}
\begin{figure} [h!]
\begin{center}
\includegraphics[width=11.0truecm]{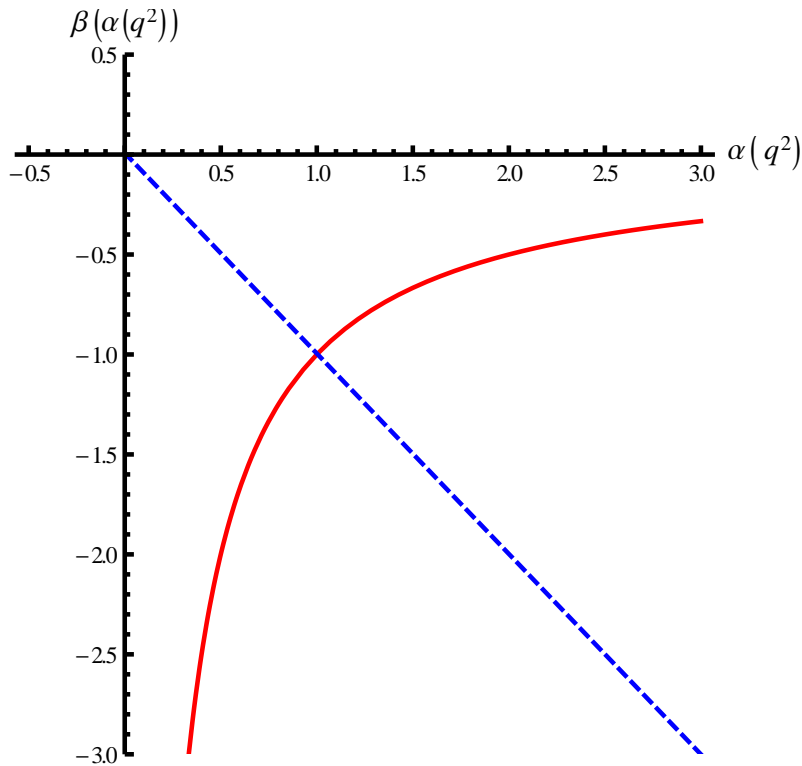}
\caption{ The straight dash line shows the $\beta$-function as a function of the effective charge $\alpha(q^2)$. The solid curve stands for it
as a function of $z$, see eq.~(C3).}
\label{fig:2}
\end{center}
\end{figure}

\hspace{2mm}

The another interesting phenomenological consequence which directly follows from the mass gap approach to QCD at large
distances ($q^2 \rightarrow 0$) is the relation between the renormalized running effective charge (B2)

\begin{equation}
\alpha(q^2; \Delta_{JW}^2) = { \Delta_{JW}^2 \over q^2}=  {1 \over z}, \quad z = {q^2 \over \Delta^2_{JW}}
\end{equation}
and the corresponding $\beta$-function. The renormalization group equation for the renormalized effective charge, which determines the corresponding
$\beta$-function, is as follows:

\begin{equation}
q^2 {d \alpha(q^2; \Delta^2_{JW}) \over dq^2} = \beta( \alpha(q^2; \Delta^2_{JW})).
\end{equation}
Then from eq.~(C1) one immediately obtains

\begin{equation}
\beta( \alpha(q^2; \Delta^2_{JW})) = - \alpha(q^2; \Delta^2_{JW}) = - {1 \over z}.
\end{equation}

Thus, one can conclude that the corresponding $\beta$-function as a function of its arguments is always in the domain of attraction, i.e., it is always negative.
So it has no IR stable fixed point. Moreover, it is scale independent as well, i.e., does not depend
on the value of the mass gap $\Delta^2_{JW}$, see Fig. 2. In other words, it posses both features required for the confining theory~\cite{2}, indeed.
In phenomenological applications the $\beta$-function itself and the ratio

\begin{equation}
{\beta( \alpha(q^2; \Delta^2_{JW})) \over \alpha(q^2; \Delta^2_{JW})} = - 1,
\end{equation}
one needs to know in order to calculate the truly NP/INP quantities, which are free from all the types of the PT contributions ("contaminations").
These are the above-mentioned YM part of the bag constant, the gluon condensate, the topological susceptibility etc.~\cite{8,45,46,47,48,49,50,51}
(and references therein).

In the previous appendix it was demonstrated that the growing effective charge (C.1) leads to the linear rising potential between heavy quarks
in complete agreement with lattice calculations. Indeed, there already exists direct lattice evidence of such kind behavior of the
full gluon propagator~\cite{52,53,54,55,56}. A NP finite-size scaling technics was used in \cite{57} to study the evolution of the running coupling
in $SU(3)$ YM lattice theory. By using the two-loop $\beta$-function it is shown to evolve according to the PT at high energies, while at low energies it
is shown to grow.

%\vfill

%\eject

%\section*{References}for example

{}


\begin{thebibliography}{}
\bibitem{1}
   H. Fritrsch, M. Gell-Mann, H. Leutwyler, Phys. Lett. B, 47 (1973) 365.
\bibitem{2}
   W. Marciano, H. Pagels, Phys. Rep. C, 36 (1978) 137.
\bibitem{3}
   M.E. Peskin, D.V. Schroeder, An Introduction to Quantum Field Theory (Addison-Wesley, 1995).
\bibitem{4}
   C. Itzykson, J.-B. Zuber, Quantum Field Theory (McGraw-Hill Book Company, 1984).
\bibitem{5}
   T. Muta, Foundations of QCD (Word Scientific, 1987).
\bibitem{6}
   N. Brambilla et al., Eur. Phys. J. C, 37 (2014) 2981. %\ arXiv:1404.3723.
\bibitem{7}
   A.S. Kronfeld, C. Quigg, Am. J. Phys., 78 (2010) 1081. %\ arXiv:1002.5032.
\bibitem{8}
   V. Gogokhia, G.G. Barnaf\"oldi, The Mass Gap and its Applications (World Scientific, 2013).
\bibitem{9}
   A. Jaffe, E. Witten, Yang\,--\,Mills Existence and Mass Gap, \\
   $http://www.claymath.org/prize-problems/, \
   http://www.arthurjaffe.com$ \ .
\bibitem{10}
   E.G. Eichtein, F.L. Feinberg, Phys. Rev. D, 10 (1974) 3254.
\bibitem{11}
   M. Baker, C. Lee, Phys. Rev. D, 15 (1977) 2201.
\bibitem{12}
   U. Bar-Gadda, Nucl. Phys. B, 163 (1980) 312.
\bibitem{13}
   R. Alkofer, L. von Smekal, Phys. Rep. C, 353 (2001) 281.
\bibitem{14}
   C.D. Roberts, A.G. Williams, Prog. Part. Nucl. Phys., 33 (1994) 477.
\bibitem{15}
   P. Maris, C.D. Roberts, Int. J. Mod. Phys. E, 12 (2003) 297.
\bibitem{16}
   J. C. Taylor, Nucl. Phys. B, 33 (1971) 436.
\bibitem{17}
   A. A. Slavnov, Sov. Jour. Theor. Math. Phys., 10 (1972) 99, 153.
\bibitem{18}
   G. 't Hooft, Nucl. Phys. B, 33 (1971) 173.
\bibitem{19}
   S.K. Kim, M. Baker, Nucl. Phys. B, 164 (1980) 152.
\bibitem{20}
   S.-H.H. Tye, E. Tomboulis, E.C. Poggio, Phys. Rev. D, 11 (1975) 2839.
\bibitem{21}
   P. Pascual, R. Tarrach, Nucl. Phys. B, 174 (1980) 123.
\bibitem{22}
   B.W. Lee, Phys. Rev. D, 9 (1974) 933.
\bibitem{23}
   S. Aoki et al., Eur. Phys. J. C, (2017) 77:112. %\ arXiv:1607.00299.
\bibitem{24}
   V.N. Gribov, J. Nyiri, Quantum Electrodynamics (Cambridge University Press, 2001).
\bibitem{25}
   V. Gogokhia, Phys. Lett. B, 618 (2005) 103.
\bibitem{26}
   G. 't Hooft, M. Veltman, Nucl. Phys. B, 44 (1972) 189.
\bibitem{27}
   G. 't Hooft, Nucl. Phys. B, 35 (1971) 167.
\bibitem{28}
   T.D. Lee, C.N. Yang, Phys. Rev., 128 (1962) 885.
\bibitem{29}
   K. Fujikawa, B.W. Lee, A.I. Sanda, Phys. Rev. D, 6 (1972) 2923.
\bibitem{30}
   B.W. Lee, J. Zinn-Justin, Phys. Rev. D, 7 (1973) 1049.
\bibitem{31}
   V. Gogokhia, G.G. Barnafoldi, Intr. Jour. Mod. Phys. A, 31 (2016) 1645027.
\bibitem{32}
   M.A. Lavrentiev, B.V. Shabat, Methods of the theory of functions of complex variable (Nauka, Moscow, 1987).
\bibitem{33}
   G.A. Korn, T.M. Korn, Mathematical Handbook (McGraw-Hill Book Company, 1968).
\bibitem{34}
   V. Gogohia, Phys. Lett. B, 584 (2004) 225.
\bibitem{35}
   G.G. Barnafoldi, V. Gogokhia, \ arXiv:1904.07748 \ [hep-ph].
\bibitem{36}
   I.M. Gelfand, G.E. Shilov, Generalized Functions, volumes I, II (Academic Press, NY, 1968).
\bibitem{37}
   G. Leibbrandt, Noncovariant Gauges (World Scientific, 1994).
\bibitem{38}
   A. Basseto, G. Nardeli, R. Soldati, Yang-Mills Theories in Algebraic Non Covariant Gauges (WS, 1991).
\bibitem{39}
   I.V. Andreev, QCD and hard processes at high energies (Moscow, Nauka, 1981).
\bibitem{40}
   M. Tanabashi, {\it et al.}, Particle Data Group, Phys. Rev. D, 98 (2018) 030001.
\bibitem{41}
   F. Wilczek, Proc. Inter. Con., QCD-20 Years Later, Vol. 1 (Aachen, June 9-13, 1992).
\bibitem{42}
   A.I. Alekseev, B.A. Arbuzov, Phys. Atom. Nucl., 61 (1998) 264.
\bibitem{43}
   K.D. Born et al., Phys. Lett. B, 329 (1994) 325.
\bibitem{44}
   V.M. Miller et al., Phys. Lett. B, 335 (1994) 71.
\bibitem{45}
   V. Gogohia, Gy. Kluge, Phys. Rev. D, 62 (2000) 076008.
\bibitem{46}
   G.G. Barnafoldi, V. Gogokhia, J. Phys. G, 37 (2110) 025003.
\bibitem{47}
   G.Gabadadze, M.A. Shifman, Int. J. Mod. Phys. A, 17 (2002) 3689. %hep-ph/0206123.
\bibitem{48}
   V.A. Novikov, M.A. Shifman, A.I. Vainshtein, V.I. Zakharov, Nucl. Phys. B, 191 (1981) 301.
\bibitem{49}
   I. Halperin, A. Zhitnitsky, Nucl. Phys. B, 539 (1999) 166.
\bibitem{50}
   E. Witten, Nucl. Phys. B, 156 (1979) 269.
\bibitem{51}
   G. Veneziano, Nucl. Phys. B, 159 (1979) 213.
\bibitem{52}
   C. Michael, Nucl. Phys. B, 42 (1995) 147.
\bibitem{53}
   G.Damm, V. Kerler, V.K. Mitrjushkin, Phys. Lett. B, 443 (1998) 88.
\bibitem{54}
   G. Burgio, F. Di Renzo, C. Parrinello, C. Pittori, Nucl. Phys. B, 73 (1999) 623.
\bibitem{55}
   G. Burgio, F. Di Renzo, G. Marchesini, E. Onofri, Nucl. Phys. B, 63(A-C) (1998) 805.
\bibitem{56}
   A.V. Nesterenko, Phys. Rev. D, 64 (2001) 116009.
\bibitem{57}
   M.Lusher et al., Nucl. Phys. B, 413 (1994) 481.
\bibitem{58}
   F. Wilczek, Phys. Today, Nov. issue (1999) 11; ibid, Jan. issue (2000) 13.
\bibitem{59}
   V. Rubakov, Classical Theory of Gauge Fields (Prinseton University Press, 2002).
\bibitem{60}
   R. Feynman, Nucl. Phys. B, 188 (1981) 479.
\bibitem{61}
   D.J. Gross, F. Wilczek, Phys. Rev. Lett., 30 (1973) 1343.
\bibitem{62}
   H.D. Politzer, Phys. Rev. Lett., 30 (1973) 1346.
\bibitem{63}
   S. Coleman, E. Weinberg, Phys. Rev. D, 7 (1973) 1888.
\bibitem{64}
   D.J. Gross, A. Neveu, Phys. Rev. D, 10 (1974) 3235.
\bibitem{65}
   M. Schottenloher, A Mathematical Introduction to Conformal Field Theory (Springer-Verlag, Berlin, Heidelberg, 1997).
\bibitem{66}
   N.F. Robertson, J.L. Jacobson, H. Saleur, arXiv:2012.07757  \ [hep-th].
\bibitem{67}
   V. Gogohia, Phys. Lett. B, 611 (2005) 129.
\bibitem{68}
   V. Gogohia, Phys. Lett. B, 468 (1999) 279.
\bibitem{69}
   S. Weinberg, The quantum theory of fields (Cambridge University Press, 1995).
\bibitem{70}
   V. Gogohia, Gy. Kluge, H. Toki, T. Sakai, Phys. Lett. B, 453 (1999) 281.
\bibitem{71}
   V. Gogohia, H. Toki, Phys. Lett. B, 466 (1999) 305.
\bibitem{72}
   V. Gogohia, Phys. Lett. B, 485 (2000) 162.
\bibitem{73}
   V. Gogohia, H. Toki, T. Sakai, Gy. Kluge, Int. J. Mod. Phys. A, 15 (2000) 45.
\end{thebibliography}
\end{document}